\documentclass{aa}  

\usepackage{graphicx}
\usepackage{txfonts}
\usepackage{lipsum}
\usepackage{subcaption}         
\usepackage{lscape}             
\usepackage{placeins}           
\usepackage{float}      
\usepackage{booktabs}    
\usepackage{siunitx}     
\usepackage{multirow}    
\usepackage{amsmath}     
\usepackage{makecell}   
\usepackage{array}     
\usepackage{booktabs}
\usepackage{fancyhdr}
\usepackage{hyperref} 

\begin{document}

   \title{CORN -- Chronometers of Relic Nature}
   \subtitle{I: The first estimate of the expansion rate of the Universe using compact relic galaxies 
   }


%

   \author{Krzysztof~Lisiecki\inst{\ref{ncbj}}
        \and Nicola~Principi Cavaterra\inst{\ref{ncbj}}
        \and Agnieszka~Pollo\inst{\ref{ncbj}, \ref{uj}}
        \and Chiara~Spiniello\inst{\ref{CS}, \ref{CR}}
        \and Laura~Hunt\inst{\ref{LH}}
        \and Charlie~Rosen\inst{\ref{CR}}
        \and Marek~Biesiada\inst{\ref{ncbj}}
        \and Darko~Donevski\inst{\ref{ncbj}, \ref{DD}}
        \and Patryk~Matera\inst{\ref{ncbj}}
        \and Giuliano~Lorenzon\inst{\ref{ncbj}}
        \and Katarzyna~Małek\inst{\ref{ncbj}}
        \and John~Mills\inst{\ref{JM}}
        }

   \institute{
   National Centre for Nuclear Research, Pasteura 7, 093, Warsaw, Poland; \email{krzysztof.lisiecki@ncbj.gov.pl}\label{ncbj} \and
   Astronomical Observatory of the Faculty of Physics, Astronomy and Applied Computer Science, Jagiellonian University, ul. Orla 171, 30-244 Kraków, Poland\label{uj} \and
   European Southern Observatory, Karl-Schwarzschild-Straße 2, 85748, Garching, Germany\label{CS} \and
   Sub-Dep. of Astrophysics, Dep. of Physics, University of Oxford, Denys Wilkinson Building, Keble Road, Oxford OX1 3RH, UK\label{CR}\and
   Open University, Milton Keynes, MK7 6AA, UK\label{LH}\and
   SISSA, Via Bonomea 265, 34136 Trieste, Italy\label{DD}\and
   Department of Physics, University of Warwick, Gibbet Hill Road, Coventry CV4 7AL, UK\label{JM}
   }

   \date{Received XXX XXX, 2026}

\abstract{
Measuring the expansion rate of the Universe is a central challenge in cosmology, particularly due to the persistent statistically significant discrepancy between the measurements of the Hubble constant, $H_0$, obtained using early- and late-time probes, commonly known as the Hubble tension. 
Independent and robust methods are therefore essential to validate existing measurements and assess potential systematic effects.
}
{
We employ the cosmic chronometer (CC) approach, which uses the differential age evolution of quiescent galaxies to measure the Hubble parameter $H(z)$ without assuming an underlying cosmological model. 
In contrast to previous CC studies, we restrict our analysis to relics -- ultra compact massive galaxies (UCMGs) hosting the oldest stellar populations -- thereby minimising uncertainties related to star formation history and merger history.
}
{
We select a sample of 189 relic galaxies in the redshift range $0.07\leq z\leq0.22$ from the E-INSPIRE UCMGs catalogue. 
Using the $D_n4000$ spectral index of relic galaxies and its redshift evolution, combined with MILES stellar population synthesis models, we derive the differential age relation required to infer $H(z)$.
We account not only for metallicity effects but also for the impact of $\alpha$-element enhancement, which has not been explicitly propagated into the systematic uncertainty budget of the $D_n4000$ cosmic chronometer method.
}
{
We obtain an independent measurement of $H(z=0.15) = 85\pm53  \textrm{ km\,s}^{-1}\,\textrm{Mpc}^{-1}$. 
Although the total uncertainty remains large and dominated by statistical limitations, the systematic component is 
13\% (8.7\% when $\alpha$-enrichment is not propagated), among the tightest estimates in current CC studies. 
We find that $\alpha$-enrichment plays a major role in the systematic error budget, particularly in the high metallicity regime, and must be properly accounted for in future analyses.
}
{
We demonstrate the proof of concept that relic galaxies provide a promising pathway to reduce systematic uncertainties in the CC method. The statistical uncertainties, which dominate the error budget, can be significantly reduced by the increase of data volume expected during the next years with \textit{Euclid}, Vera C. Rubin Observatory, DESI, and 4MOST.
}
 

   \maketitle

\section{Introduction}
Cosmology has undergone rapid development over the past decades. 
Since the early 2000s, a wealth of observational evidence has established the $\Lambda$CDM model as the standard cosmological framework \citep[][]{Perivolaropoulos22}.
Its main components are the cosmological constant $\Lambda$, responsible for the accelerated expansion of the Universe, and cold dark matter (CDM), which, together with baryons, accounts for the gravitating matter content \citep[][]{Hinshaw13, Alam17, Bertone18, planck16, Planck}.
Despite its success, the standard model has recently been challenged by several tensions.
Most notably, the value of the Hubble constant, $H_0$, inferred from low-redshift distance ladder measurements (e.g. Cepheids and type Ia supernovae) differs at the $\sim 4\sigma$ level from the value derived from CMB observations.
Resolving this discrepancy, commonly referred to as Hubble tension (see, e.g. \citealp{Verde, DiValentino, Hu_H0}) requires independent cosmological probes, which can help to distinguish between unaccounted systematic effects and possible extensions to the standard model.
In recent years, several alternative methods have been developed to complement traditional probes such as CMB, baryon acoustic oscillations, and supernovae (see \citealp{Bernal16, Shah21, Morescoreview} for reviews on the topic).

The cosmic chronometer (CC) method, first proposed by \cite{JimenezFirst}, is based on the differential age evolution of the so-called "quiescent" galaxies. 
This approach does not require the assumption of a specific cosmological model, relying only on the validity of the Friedmann-Lemaître-Robertson-Walker metric.
Under this assumption, the Hubble parameter can be expressed as:
\begin{equation}\label{eq:hz_dzdt}
H(z) \equiv \frac{\dot{a}}{a} = -\frac{1}{1+z}\frac{\mbox{d}z}{\mbox{d}t},
\end{equation}
where $z$ is the redshift and $a$ is the scale factor. Thus, by measuring the age evolution of a synchronised population of tracers, one can estimate $\mbox{d}z/\mbox{d}t$ and directly constrain $H(z)$.
The optimal tracers for the CC method are old and quiescent galaxies due to homogeneous and synchronised formation histories \citep{MorescoCC1}.

However, determining galaxy ages is subject to significant uncertainties.
Degeneracies in stellar population modelling and the presence of younger stellar components can bias and/or reduce the precision of age determination.
One of the most important degeneracies is the age-metallicity, in which older, metal-poor populations can mimic the spectral features of younger, metal-rich ones \citep[][]{Worthey94, Kaviraj07}.
Breaking this degeneracy requires independent constraints on stellar metallicity ($Z$), introducing additional modelling assumptions and systematic uncertainties.
Although the impact of $Z$ has been extensively studied in the CC context by establishing its influence on the uncertainty, controlling these systematics remains a major challenge \citep[][]{Moresco16, Loubser25+, Tomasetti25}.
An additional and largely unexplored source of systematic uncertainty arises from variations in detailed chemical abundance patterns, in particular from the enhancement of $\alpha$-elements.
The latter are mostly produced on short timescales in core-collapse supernovae, while the iron-peak elements released by type Ia supernovae are produced on longer timescales \citep[over Gyrs;][]{Thomas99, Maoz10, Vazdekis15}. 
Thus, their relative abundance ([$\alpha$/Fe] ratio) is a proxy of the star formation timescale: higher values correspond to rapid and early star formation \citep[][]{Thomas05}.
Massive early-type galaxies, commonly used as CC tracers, are known to exhibit super-solar [$\alpha$/Fe] \citep[][]{Walcher15}.
Since stellar population synthesis (SPS) models are sensitive to abundance patterns, neglecting variations in [$\alpha$/Fe] can bias age estimates and propagate directly into $H(z)$ assessments.
Previous studies have generally argued that the influence of $\alpha$-enhancement can be neglected in the systematic uncertainty budget \citep{Moresco12,Moresco18}, although more recent works have measured [$\alpha$/Fe] on individual galaxies using variable-abundance stellar population models \citep{Borghi22,Alvarez25}. 
Here, we explicitly propagate the uncertainty associated with [$\alpha$/Fe] into the $H(z)$ measurement based on the $D_n4000$ method and assess its contribution to the systematic error budget.

So far, CC studies have primarily relied on massive, passively evolving elliptical galaxies (early type galaxies, ETGs).
Although these systems form the bulk of their stellar mass in short early bursts, their subsequent evolution is often driven by mergers \citep[e.g.][]{naab09, oser10, zolotov15, flores-freitas21}.
Such processes increase the size of the galaxy and can introduce younger stellar populations \citep{vandokkum10}, and/or regenerate star formation \citep[][]{Chauke19, Szpila25}, thus affecting the assumptions required for ideal CC tracers.
One way to address this is to isolate ETGs contaminated by younger stellar populations following the guidelines highlighted in \cite{Moresco18}. 
However, in this proof-of-concept paper, we propose an alternative approach based on a rare but powerful category of galaxies, generally referred to as  massive relics \citep{Trujillo14, ferre-mateu17, Spiniello21, Spiniello24, Comeron+23, Mills25, Pascalau26}

Relics are a sub-sample of massive and passive galaxies, classically used as CCs. 
They are the local descendants of high-z red nuggets \citep[e.g.][]{Damjanov+11, Lisiecki23}, i.e. ultra-compact (R$_{\textrm{eff}}<2$ kpc) and massive (M$_{\star}>6\times10^{10}$M$_{\odot}$) galaxies (UCMGs) that have avoided significant merger activity and remain extremely compact at low redshift. 
Their stellar populations are extremely old and formed in the first phases of the cosmic history \citep[][]{Trujillo14, ferre-mateu17, Spiniello21, Spiniello24}. 
Relics are characterised by extreme star formation histories (SFHs): they formed the almost totality of their stellar mass in a quick and violent burst of star formation at very high redshift, and then evolved passively and undisturbed, without forming any new stars nor accreting anything throughout their cosmic life.

In this work, we investigate relic galaxies from the E-INSPIRE catalogue ($z\sim0.15$) as CCs, motivated by their potential to provide a homogeneous population with minimal evolutionary contamination.
Additionally, we further explore the measured enhancement of the $\alpha$-elements as a source of systematic uncertainty in the measurement of $H(z)$.

\section{Data}

We use the largest available spectroscopic sample of UCMGs, from the E-INSPIRE study \citep{Mills25, Rosen+26}.
These objects were selected from the Sloan Digital Sky Survey Data Release 18 \citep[SDSS;][]{Almeida23} with the following criteria:
\begin{itemize}
    \item $\Sigma_{1.5} > 10.5$, where $\Sigma_{1.5} = \log(M_\star/R_e^{1.5})$ is the compactness parameter introduced by \citet{Barro13} and the threshold adopted from \citet{Baldry21}, with stellar masses derived from the GALEX-SDSS-WISE Legacy Catalogue \citep[GSWLC-2;][]{Salim18};
    \item $M_\star > 6 \times 10^{10}\,M_\odot$ to ensure only massive sources;
    \item $R_e < 2$ kpc to ensure only compact sources.
\end{itemize}
Star-forming systems were excluded using a colour cut ($g-i > 1.2$) and SDSS quality flags, ensuring a sample dominated by old quiescent stellar populations.
A detailed description of the selection is provided in \cite{Mills25}.
While this selection alone may retain dusty star-forming galaxies, such contaminants are subsequently rejected through the full spectral fitting.

The E-INSPIRE sample contains 430 UCMGs in the redshift range $0.01 < z < 0.3$.
We present the distribution of stellar mass (M$_\star$) versus $z$ in Figure~\ref{fig:sample}.
The spectra from SDSS cover 3800–9200\AA, while the signal to noise ratio for UCMGs selected by \cite{Mills25} ranges from 10-50.
Stellar population properties were derived using {\tt PENALISED PIXEL-FITTING} \citep[{\tt pPXF}; ][]{Cappellari04, Cappellari17, Cappellari23} on the SDSS spectra with E-MILES SPS models \citep{Vazdekis16}.
For each galaxy, this provides estimates of stellar $Z$, [$\alpha$/Fe], and the SFH.
In this work, we adopt the updated measurements from \cite{Rosen+26}, which include models that extend to the highest metallicities ($[Z/\mbox{H}] = 0.4$).

\subsection{Degree of relicness}\label{sec:dor}
To quantify how much an UCMG changed during its lifetime, a concept of "degree of relicness" (DoR, hereafter) was introduced by \cite{ferre-mateu17}.  
This was then operationally defined in \cite{Spiniello24} as a dimensionless number that combines the formation and mass assembly timescales as follows:
\begin{equation}\label{Eq:dor}
    \textrm{DoR} = \left[ f_{\textrm{M}^\star_{t\textrm{BB=3}}} + \frac{0.5\textrm{Gyr}}{T_{75}} + \frac{0.7\textrm{Gyr} + (T_\textrm{Uni} - T_\textrm{fin})}{T_\textrm{Uni}}\right] \times \frac{1}{3},
\end{equation}
where $f_{\textrm{M}^\star_{t\textrm{BB=3}}}$ is the fraction of the mass assembled within 3~Gyr after the Big Bang, $T_{75}$ is the cosmic time at which 75 percent of the M$_\star$ was in place, $T_\textrm{fin}$ is the cosmic time at which 99.8 percent of the M$_\star$ was in place (not 100\% due to residual star formation), and $T_\textrm{Uni}$ is the age of the Universe at observed redshift.
The normalisation factors follow recent work motivated by observations of extremely early-formed massive galaxies, prototypical relics \citep{labbe23}. 
The higher the DoR, the closer the galaxy is to a true relic, i.e. DoR$=1$ is the prototypical relic.
Each UCMG within the E-INSPIRE sample has an estimated DoR according to the {\tt pPXF} analysis provided by the original catalogue.

Although the computation of the DoR requires an assumed cosmological model to convert stellar ages into cosmic time, it primarily parameterises the shape of the reconstructed SFH rather than its absolute time scale. 
Nevertheless, the adopted cosmology can influence the recovered DoR through the age constraints imposed during the {\tt pPXF} fitting. 
To quantify this effect, we repeated the analysis using several commonly adopted cosmological models (Appendix~\ref{app:cosmodep}). 
We find that the resulting changes in the DoR values have a negligible impact on the final selection and on the derived $H(z)$ measurement. 
All DoR values presented in the main part of this work are computed assuming the fiducial $\Lambda$CDM cosmology of \citet{Planck}.

To ensure the robustness of spectral fitting, we estimate its uncertainty using four {\tt pPXF} runs performed by the E-INSPIRE team, in addition to individual DoR values. 
In particular, in \cite{Mills25} and \cite{Rosen+26}, a value for [Mg/Fe] is initially computed using line-index measurements and used as a proxy for the abundance of [$\alpha$/Fe]. 
Then, four different full-spectral fitting runs were executed: 
\begin{itemize}
    \item the best individual fit using as input only Single stellar populations (SSP) models with the corresponding [$\alpha$/Fe] (DoR$_\textrm{unreg}$);
    \item an individual fit with SSP models having [$\alpha$/Fe] abundance higher by 0.1 with respect to the SSP-like estimate (DoR$_\textrm{plus}$);
    \item an individual fit with SSP models having [$\alpha$/Fe] abundance lower by 0.1 with respect to the SSP-like estimate (DoR$_\textrm{minus}$);
    \item the resulting fit obtained by bootstrapping the pPXF run in 9 iterations, each time resampling the spectra, but always assuming the right SSP-like [$\alpha$/Fe] abundance for the input models (DoR$_\textrm{reg}$).
\end{itemize}
Each of the above mentioned runs results in a slightly different DoR estimation \citep[details in ][]{Mills25}. 
We define the uncertainty on DoR, $\sigma_\textrm{DoR}$, as the scaled median absolute deviation (MAD, {scaled~MAD~$\equiv1.4826\times$MAD).

\subsection{Final sample}\label{sec:finalsample}
We remind the reader that limiting ourselves to UCMGs from the E-INSPIRE sample, by construction we limit our study to ultra compact morphology. 
This minimises the potential risk of mergers and inflow, and thus the risk of combining two progenitor galaxies into composite stellar population.
Moreover, to ensure that we are only selecting red and dead galaxies, populated in great majority by very old stars, we exclude 30\% of galaxies with the lowest DoR, which in our case translates into $\textrm{DoR}\gtrsim 0.4$\footnote{
Objects with a DoR$<0.4$ have formed less than 80\% of their stellar mass during the first phase of the two-phase formation scenario (\citealp{Spiniello24}, Table 3), meeting the operative definition of 'relic' given in \cite{Spiniello21}. 
}. 
To ensure the robustness of the DoR estimates, we additionally exclude galaxies with $\sigma_{\rm DoR}>0.2$, corresponding to approximately the upper 20\% of the $\sigma_{\rm DoR}$ distribution.
By allowing sources with poorly constrained DoR, we are losing the most important constraint on the age.
We verified that relaxation of this threshold leads to the diluting of the sample and the introduction of contaminates.
Conversely, adopting a more restrictive threshold does not significantly improve the quality of the fit, while unnecessarily reducing the sample size.
We therefore adopt $\sigma_{\rm DoR}\leq0.2$ as the best compromise between robust DoR estimates and sufficient statistical power.

To establish a homogeneous and statistically robust sample, we apply additional selection cuts.
We restrict the sample to $0.07 \leq z \leq 0.22$ and $\textrm{M}_\star \geq 10^{10.8}\,\textrm{M}_\odot$.
The lower redshift limit minimises the volume effects and the scarcity of massive systems at very low redshift, whereas the upper limit minimises the incompleteness issue of the UCMG selection and ensures enough sources in the whole redshift range.
The mass cut is applied to maintain consistency with the high-mass regime, where CCs should be reliably identified.
We note that, despite these cuts, the sample may still suffer from residual incompleteness. However, adopting more restrictive selection criteria would significantly reduce the sample size and compromise the statistical robustness of the CC analysis.

Thus, the final sample consists of 189 galaxies with $0.07 \leq z \leq 0.22$, $\textrm{M}_\star \geq 10^{10.8}\,\textrm{M}_\odot$, $\sigma_\textrm{DoR}\leq 0.2$, and DoR $\gtrsim 0.4$.
The distributions of $z$ and M$_\star$ of the final sample are presented in Figure~\ref{fig:sample}.

\begin{figure}[h]
\centering
\includegraphics[width = 0.48\textwidth]{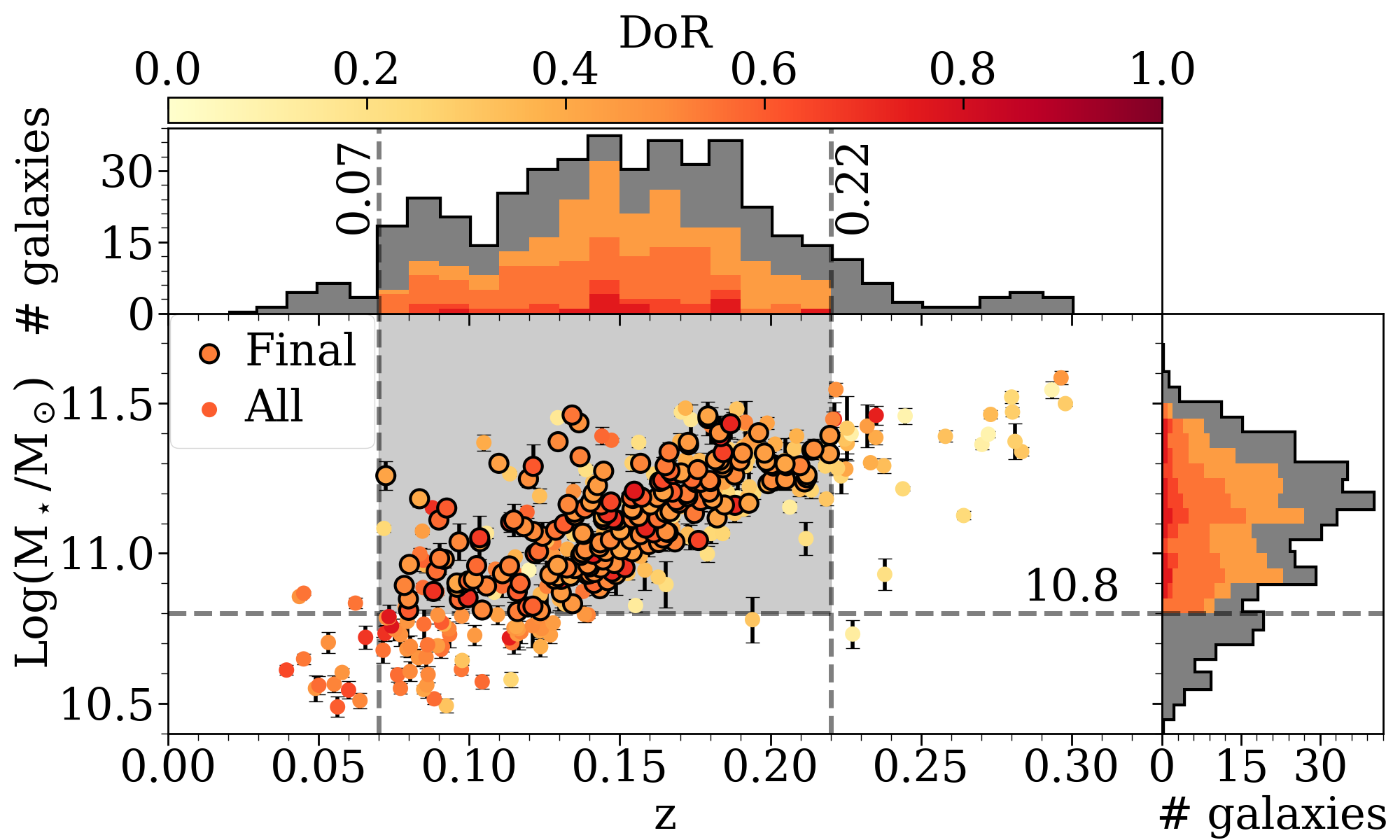}
\caption{M$_\star$ as a function of redshift for E-INSPIRE galaxies, colour-coded by their DoR.
The grey dashed lines show the cuts performed, and the shaded region show the cuts-limited region. 
Galaxies in the final sample are highlighted with a black solid contour circle. 
Histograms on the top- and right-side show the distribution of M$_\star$ and redshift covered by the galaxies. 
The coloured histograms encode the DoR intervals within the final sample distributions.
}
\label{fig:sample}
\end{figure}

It is worth emphasizing that our selection is not directly comparable to those adopted in traditional CC studies \citep[e.g.,][]{Moresco18,Alvarez25}. 
Relic galaxies represent a highly restrictive subset of the massive and passive galaxy population. 
To quantify the overlap between the two approaches, we applied a standard CC selection based on commonly adopted criteria, including $UVJ$ colours, emission-line diagnostics, H$\delta$ absorption, the H:K ratio, and stellar velocity dispersion (see Appendix~\ref{app:traditonal}). 
We find that only 30\% of our final relic sample satisfies all classical CC criteria, although this fraction increases to 87\% when the velocity dispersion requirement is omitted. 
Therefore, while most relic galaxies would also be classified as classical CCs, standard CC selections may exclude a substantial fraction of relic systems.

An additional distinction lies in the size of the resulting samples.
Using SDSS data (the same parent sample as E-INSPIRE project), \citet{Alvarez25} identified a CC sample that is approximately 500 times larger than ours. 
Even after accounting for the broader redshift range considered in their study, their sample remains about two orders of magnitude larger.
This suggests that our relic-based selection targets a substantially more restrictive subset of the passive galaxy population rather than a direct equivalent of the standard CC selection.

\section{Relics as cosmic chronometers}\label{sec:sfh}

As stressed in the introduction, the use of red and dead massive galaxies as CCs has proven to be successful, since the uncertainties associated with prolonged or recent star formation episodes can be significantly minimised. 
However, although these galaxies are predominantly quiescent and their star formation timescales are known to be inversely correlated with stellar mass \citep{Thomas05}, massive early-type galaxies (ETGs) are also widely believed to have assembled through hierarchical merging and accretion processes, including both major and minor mergers \citep{Tomre77, DeLucia06, naab09, oser10, vandokkum10}. 
Relics, instead, completely omitted the merger and gas accretion phase and had a much more extreme SFH. 
To support the SFH argument, we examine the mass assembly histories of galaxies in our final sample, and compare it to other samples of normal-size, equally massive and passive galaxies and to simulations. 
In Figure~\ref{fig:massFraction}, we show the fraction of stellar mass assembled within $\sim$3 Gyr after the Big Bang. 
The median value ($\pm$ scaled MAD) for the final sample is $f_{\mathrm{M}^\star_{t_{\mathrm{BB}}=3}} = 0.95 \pm 0.05$, compared to $0.92 \pm 0.08$ for the UCMGs from the E-INSPIRE sample. 
Although the median values are similar, the final sample exhibits a significantly reduced scatter and a lower fraction of late-forming outliers.

We compare these results with massive ($M_\star > 10^{11} M_\odot$) quiescent  (according to rest-frame UVJ colours) galaxies from the LEGA-C survey \citep{vanderWel16}, analysed by \cite{Kaushal24}. 
Their SFHs were reconstructed using two independent SED-fitting codes, {\tt PROSPECTOR} \citep[][]{Johnson21} and {\tt BagPipes} \citep[][]{Carnall18}. 
The median mass fraction assembled within 3 Gyr after the Big Bang is $\sim$0.68 and $\sim$0.38 for {\tt PROSPECTOR} and {\tt BagPipes}, respectively. 
This indicates that typical massive quiescent galaxies assemble a substantially smaller fraction of their stellar mass at early times compared to our relic sample. 
Importantly, the LEGA-C galaxies are observed at $0.6 < z < 1$, and therefore have had less time for late-time mass assembly than local galaxies. 
Their early mass fractions can thus be considered upper limits for the general quiescent-galaxies population at $z \sim 0$, further emphasising the extreme nature of the relic sample.

Finally, we compare our results with the SIMBA cosmological hydrodynamical simulation \citep{dave19}. 
We select massive ($M_\star > 10^{11} M_\odot$) passive galaxies (specific SFR $< 10^{-10.8}\,\mathrm{yr}^{-1}$) in snapshots spanning $0.05 < z \leq 0.23$, and track their progenitors back to $z \sim 2$. 
The $f_{\textrm{M}^\star_{t\textrm{BB=3}}}$ decreases from $\sim$0.49 at $z=0.23$ to $\sim$0.39 at $z=0.05$, in good agreement with observational results from LEGA-C.

The consistently lower early mass fractions found in both observations and simulations indicate that typical quiescent galaxies have more extended and heterogeneous SFHs. 
In contrast, the high and narrowly distributed early mass fractions in our sample imply highly synchronised formation histories, reducing the impact of extended or secondary star formation episodes on the $D_n4000$--age relation. 
This supports the assumption that SFH-related systematics are subdominant for relic-selected CCs.

\begin{figure}[]
\centering
\includegraphics[width = 0.48\textwidth]{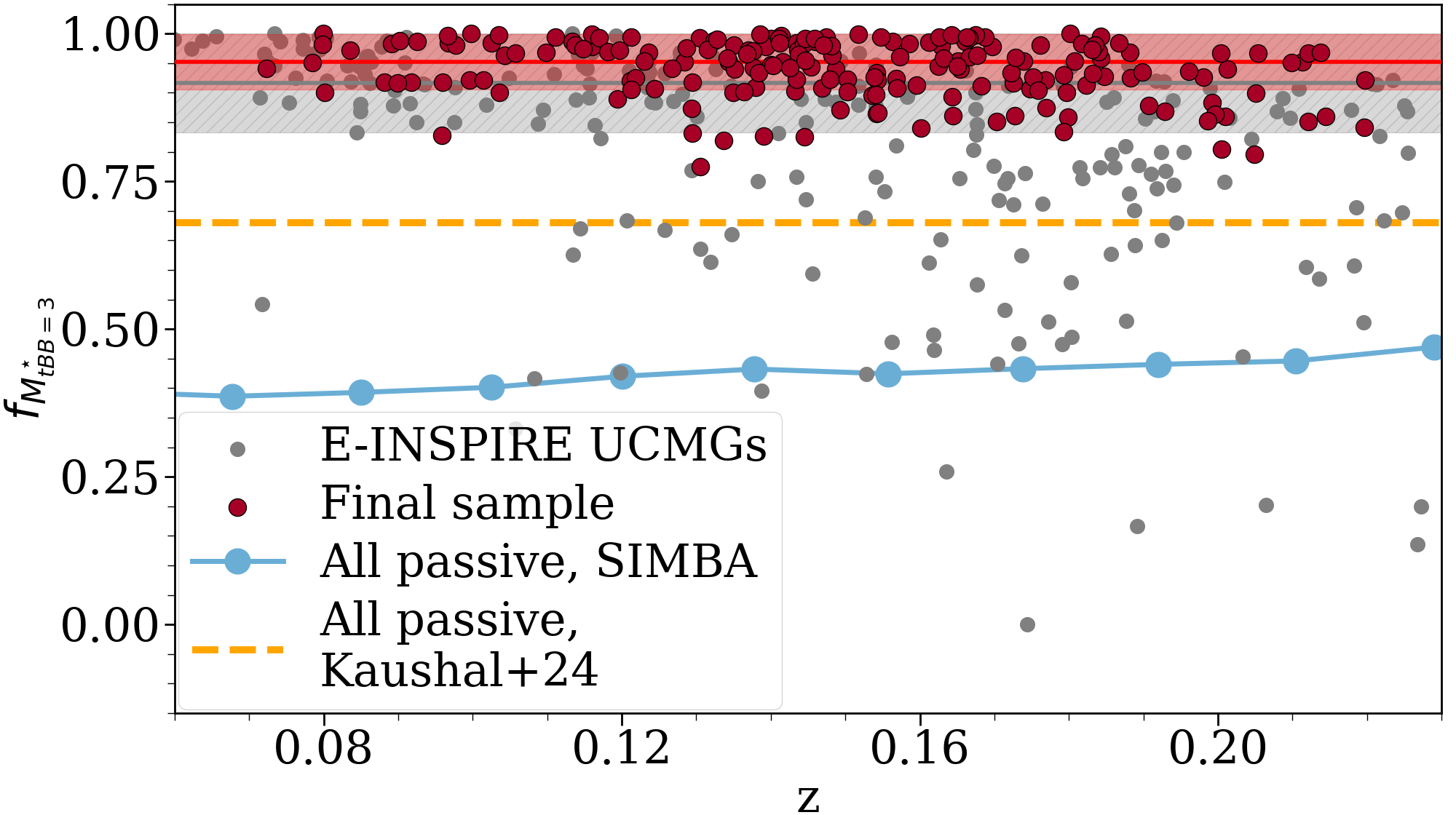}
\caption{Mass fraction assembled 3 Gyr after the Big Bang as a  function of redshift. UCMGs from E-INSPIRE are marked as gray points, while the final sample used in this paper is marked as red points. The solid lines show the median in each sample, while the shaded region represent the MAD. The orange dashed line presents the median value for all quiescent and massive galaxies by \cite{Kaushal24}.  Finally, massive and quiescent galaxies from the SIMBA simulation are shown in blue.  }
\label{fig:massFraction}
\end{figure}

\section{Methodology}
To estimate $H(z)$ (Equation~\ref{eq:hz_dzdt}), one must measure the differential age $\mathrm{d}t$ corresponding to a redshift interval $\mathrm{d}z$.
Since the 4000~$\AA$ break ($D_n4000$) correlates with the stellar population age \citep[e.g.][]{Poggianti97, Moresco11, Moresco12}, we assume the following linear relation:
\begin{equation}\label{eq:d4000slope}
    D_n4000 = A (Z, [\alpha\mbox{/Fe], M}) \times \mbox{age} + B,
\end{equation}
where $A (Z, [\alpha\mbox{/Fe], M})$ is the slope of the relation of $D_n4000$ - age, and $B$ is normalisation.
The slope (hereafter $A$) depends on the metallicity $Z$, $\alpha$-enhancement [$\alpha$/Fe], and the model M (i.e. SFH, SPS, SL, IMF, young stellar populations).
Combining Equations~\ref{eq:hz_dzdt} and \ref{eq:d4000slope} yields:
\begin{equation}\label{eq:Hz_dzdD}
    H(z) \equiv \frac{\dot{a}}{a} = -\frac{1}{1+z} A\frac{\mbox{d}z}{\mbox{d}D_n4000}.
\end{equation}
The calibration parameter $A$ absorbs the model-dependent systematic uncertainties \citep{Moresco16}, while the observational and statistical uncertainties are encoded in $\mathrm{d}z/\mathrm{d}D_n4000$.

In the following sections, we describe the estimation of $D_n4000$ for galaxies in our sample and our quantification of the $A$ parameter and its relation to stellar parameters (e.g. $Z$ and [$\alpha/$Fe]).

\subsection{The $D_n4000$ break}\label{sec:d4000}
We adopt the narrow definition of the $D_n4000$ break from \cite{Balogh99}, which is less sensitive to dust reddening than the original definition by \cite{Bruzual83}.
The $D_n4000$ index is defined it as the ratio between the continuum flux densities in a red band (${\lambda_1^\textrm{red}} = 4000,  {\lambda_2^\textrm{red}} =4100\,\AA$) and a blue band ($\lambda_1^\textrm{blue} =3850, \lambda_2^\textrm{blue}=3950\,\AA$):
\begin{equation}\label{eq:D4000}
    D_n4000 = \frac{(\lambda_2^\textrm{blue} - \lambda_1^\textrm{blue}) \int^{\lambda_2^\textrm{red}}_{\lambda_1^\textrm{red}} F_\nu d\lambda}{(\lambda_2^\textrm{red} - \lambda_1^\textrm{red}) \int^{\lambda_2^\textrm{blue}}_{\lambda_1^\textrm{blue}} F_\nu d\lambda}.
\end{equation}
The spectral regions used to calculate $D_n4000$ are highlighted in red in Figure~\ref{fig:sepctraExample}, which shows an exemplary relic spectrum and an old (13.5 Gyr) SSP template in the relevant spectral region from MILES.

\begin{figure}[t]
\centering
\includegraphics[width = 0.48\textwidth]{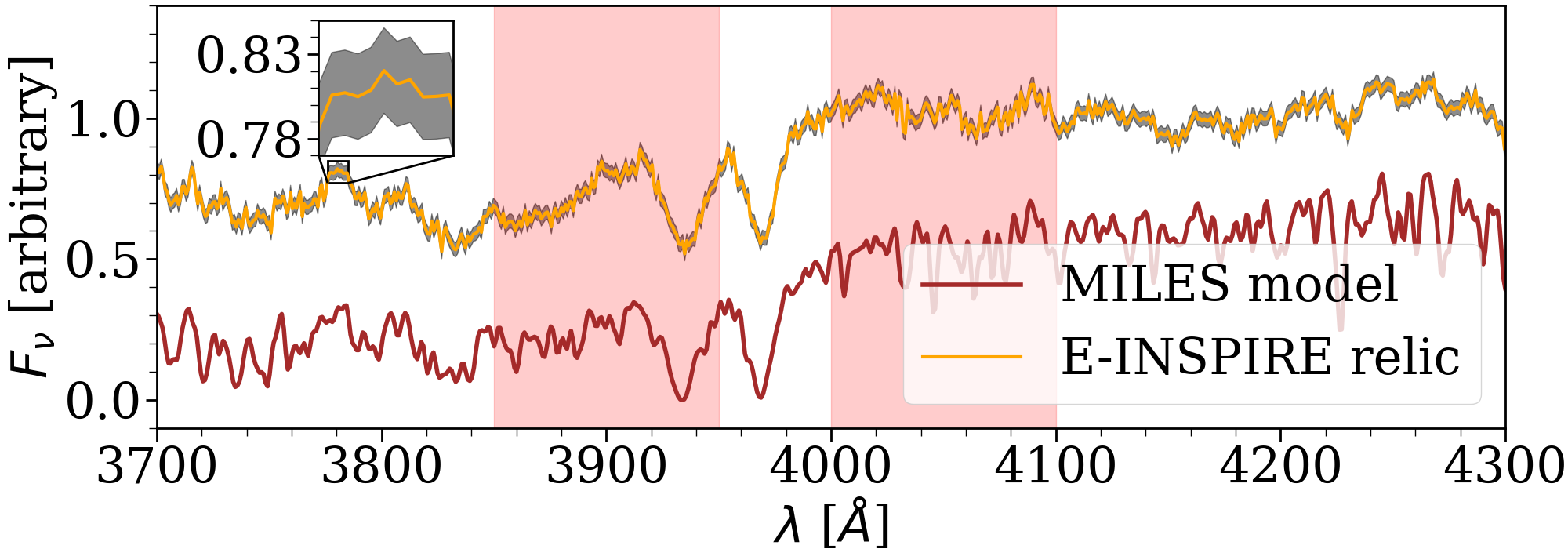}
\caption{Comparison between an exemplary relic spectrum (upper, yellow) and an old (13.5 Gyr, $Z=0.0198$) MILES SPS model (lower, red), both in restrframe wavelength. The red shaded area shows the range of blue and red bandpasses used for the estimation of $D_n4000$ (see Equation~\ref{eq:D4000}). The grey shaded region shows the uncertainties of observed spectrum.
}
\label{fig:sepctraExample}
\end{figure}

To estimate uncertainties of $D_n4000$, we use the well established algorithm {\tt DER\_SNR}\footnote{\url{https://www.stecf.org/software/ASTROsoft/DER_SNR/}} \citep[][]{Stoehr08}.
It provides a robust, model-independent estimate of the signal-to-noise ratio (SNR) directly from the spectrum by assuming that the underlying signal varies smoothly compared to the high-frequency noise. 
The method computes the noise from the scaled median absolute deviation of the second finite difference of adjacent flux samples, which suppresses slowly varying spectral features while isolating pixel-to-pixel noise.  
The SNR is then defined as the ratio of the median flux to the noise estimate, making the procedure insensitive to outliers, spectral lines, and continuum shape, and applicable even when no explicit uncertainty of the spectrum is available. 
This approach yields a reliable empirical uncertainty estimate for flux measurements in moderately well-sampled spectra.

\subsection{The $D_n4000-z$ slope}

We assume that the $D_n4000-z$ dependence is a linear relation in the relevant redshift range as in Equation~\ref{eq:d4000slope}.
To reduce the impact of the intrinsic scatter, galaxies are grouped into equally-populated  redshift bins and median values of $D_n4000$ and $z$ are computed for each bin.
The uncertainties in each bin are estimated using the scaled MAD divided by the square root of galaxies per bin. 
We fit a linear relation using total least squares, which accounts for uncertainties in both variables, providing the parametric uncertainty of the model.

To understand how much influence the size of the bin has on the final result, we perform fits with bin sizes ranging from 5 to 25 per bin. 
We found that the variation of the number of galaxies per bin does not introduce significant bias, and the values are always consistent within 1 $\sigma$ (i.e. scaled MAD). 
The result of fit with 8 galaxies per bin is presented in Figure~\ref{fig:Dn_z_slope}.

The analysis yields the following relation:
\begin{equation}
    \frac{\mbox{d}D_n4000}{\mbox{d}z}= -0.33 \pm 0.19 \textrm{ [fit] } \pm 0.07 \textrm{ [bin] },
\end{equation}
where the first term reflects the fitting uncertainty, and the second quantifies the variation with a binning scheme.
Thus, we report the total statistical uncertainty $\sigma_{stat} = 0.2$,  i.e. $61\%$ of the slope value.

\subsection{The $D_n4000$-age slope}\label{sec:D_age_slope}
To quantify how  $Z$ and [$\alpha/$Fe] affect $H(z)$, we first calibrate the relation between $D_n4000$ and age.
We use MILES SPS models with BaSTI isochrones \citep{Vazdekis15}.
This ensures consistency with the E-INSPIRE stellar population modelling\footnote{The E-MILES models used by \cite{Mills25} and \cite{Rosen+26} are only an extension to bluer and redder wavelength of the MILES SPS, but they are based on the same stars and isochrones in the optical spectral region \citep{Vazdekis16}.}
Additionally, it allows us to study the slope parameter, $A$ (Equations~\ref{eq:d4000slope},~\ref{eq:Hz_dzdD}), as a function of $Z$, and abundance of $\alpha$-elements, [$\alpha/$Fe]. 
The models are available for [$\alpha$/Fe] = 0.0 and 0.4. 
In order to have intermediate values, we linearly interpolate the spectra between those models.
We note that this interpolation assumes linearity in the spectral response with [$\alpha/$Fe], which is only an approximation. 
In the absence of a finer grid, linear interpolation provides a reasonable first-order description consistent with previous analyses \citep[][]{Spiniello21, Spiniello24, Mills25}. 
We therefore adopt a step of $\Delta[\alpha/\mathrm{Fe}] = 0.1$, as smaller increments would likely exceed the level of precision supported by the underlying models.

\begin{figure}[]
\centering
\includegraphics[width = 0.48\textwidth]{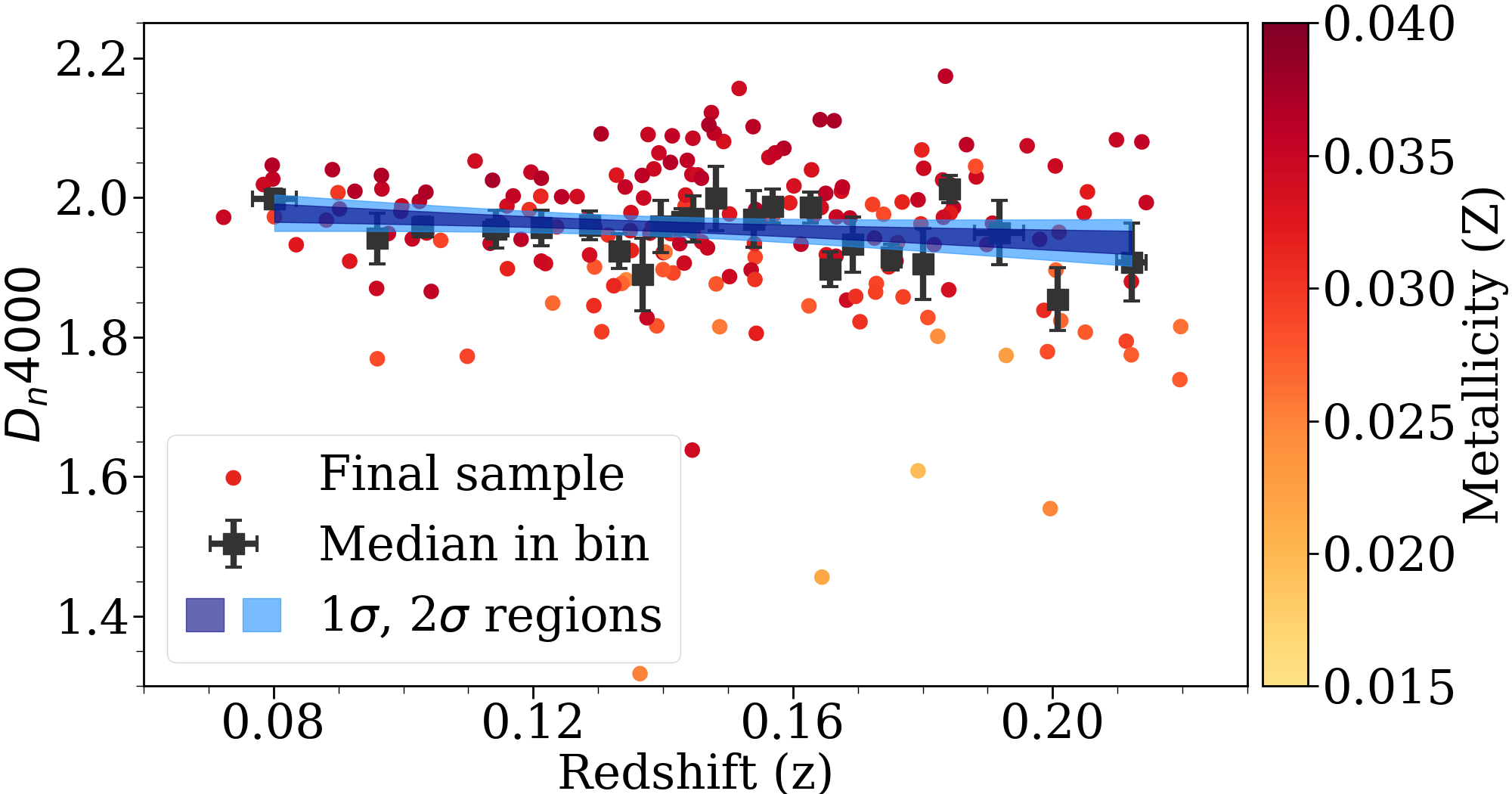}
\caption{$D_n4000$ as a function of redshift (z), colour coded by stellar $Z$. Small circles show the individual galaxies while black squares show the median within each bin built with 8 galaxies in each. The shaded areas, light and dark blue correspondingly, present 1$\sigma$ and 2$\sigma$ levels for the best linear fit.}
\label{fig:Dn_z_slope}
\end{figure}

We first parameterise the differential age normalisation $A$ according to SPS.
The available SPS models in the MILES library are spaced by 0.5 Gyr in age.  
This timescale is comparable with the star formation period in the most extreme relic galaxies (see Section~\ref{sec:dor}).
In other words, a true relic should form the majority of its stellar mass within the first 0.5 Gyr \citep[][; also see Section~\ref{sec:sfh}]{labbe23, Spiniello24}. 
Thus, we use individual models in the analysis, and we do not study the impact of composite stellar populations on the $D_n4000$-age slope.

For each synthetic MILES spectrum, $D_n4000$ is measured following the procedure described in Section~\ref{sec:d4000}.
We then group models with fixed $Z$ and fixed [$\alpha/$Fe], while allowing age to vary, and fit a linear relation between $D_n4000$ and age (Figure~\ref{fig:analysisA}, top panel, individual line).  
This provides the calibration parameter $A$ for one combination of $Z$ and [$\alpha/$Fe]. 
To assess the dependence of $A$ on $Z$, we compare the derived slopes across models with different $Z$ at fixed [$\alpha/$Fe] (Figure~\ref{fig:analysisA}, bottom panel, individual line). 
Repeating the same procedure for all available [$\alpha/$Fe] values allows us to map the dependence of $A$ on both $Z$ and [$\alpha/$Fe], and to quantify their contribution to the systematic uncertainty.

\subsection{Systematic error estimation}
Previous studies have quantified the contribution of different sources to the systematic uncertainty budget \citep[e.g.][]{Moresco18, Moresco20, Borghi22}. 
These works reported that the choice of the SPS model contributes to the relative uncertainty $\delta\sigma_{SPS}=5.5$\% (we denote $\delta\sigma$ as the relative uncertainty from now on).
The contribution due to the choice of the IMF is less than $\delta\sigma_{IMF}=1$\%, while the contribution due to the stellar library is $\delta\sigma_{SL}=7$\%, on average. 
We adopt these values for consistency.
However, we do not include systematics related to SFH.
This assumption is motivated by the relic selection, which favours galaxies with early, rapid formation, and minimal subsequent evolution (see Section~\ref{sec:sfh}).
This is further supported by the DoR selection criteria. 
\begin{figure}[]
\centering
\includegraphics[width = 0.48\textwidth]{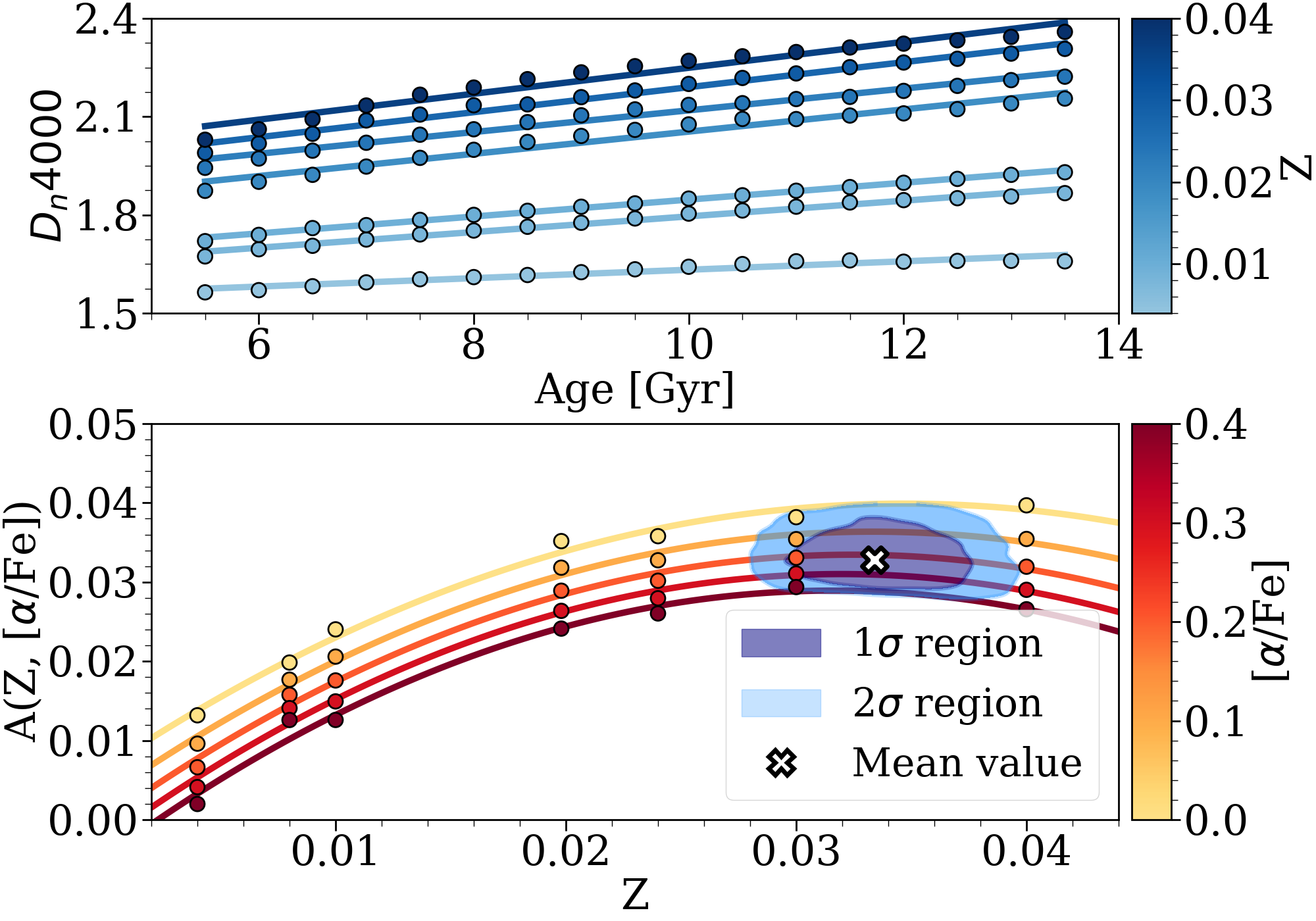}
\caption{\textit{Top}: $D_n4000$ relation as a function of stellar age and colour-coded by metallicity ($Z$).
Each point represents an SSP. Solid lines show the best linear fit for each metallicity group. \textit{Bottom:} relation of slopes $\mbox{d}D_n4000/\mbox{d}t$ as a function of stellar metallicity, colour-coded by [$\alpha$/Fe]. Each point represents one slope. Solid lines represent the best quadratic fit for each [$\alpha$/Fe] group. The blue shaded regions represent the 1 and 2$\sigma$ levels of $A$ (Equations~\ref{eq:d4000slope},~\ref{eq:Hz_dzdD}) distribution, while the white cross shows an estimated value of $A$ within our sample.}
\label{fig:analysisA}
\end{figure}

As shown in section~\ref{sec:D_age_slope} the $A =\mbox{d}D_n4000/\mbox{d}t$ depends both on $Z$ and [$\alpha$/Fe]. 
However, the systematics related to [$\alpha$/Fe] were not that extensively explored in previous studies. 
Thus, we use the measured values of [$\alpha$/Fe] of individual galaxies and the relations presented in Figure~\ref{fig:analysisA} to probe a realistic assumption of $A(Z, [\alpha/\textrm{Fe}])$.
The final sample is characterised by a median metallicity of $Z = 0.034 \pm 0.003$ and a median $\alpha$-enhancement of $[\alpha/\mathrm{Fe}] = 0.20 \pm 0.12$.
The uncertainties represent a scaled MAD.
We perform a Monte Carlo propagation of the uncertainties in $Z$ and [$\alpha$ / Fe] (see Appendix~\ref{app:Adistr}), obtaining $A(Z, [\alpha/\textrm{Fe}]) = 0.033^{+0.003}_{-0.002}$ (see the blue distribution in Figure~\ref{fig:Ahist}). 
Thus, $\delta\sigma_{Z,\alpha} =^{+9\%}_{-6\%}$.
The distribution of the results is presented in the lower panel of Figure~\ref{fig:analysisA}.

To have a direct comparison with previous CC studies, we also test the uncertainty without propagating [$\alpha$/Fe]. 
Using the median values reported above and propagating only the uncertainties on metallicities (see Appendix~\ref{app:Adistr}) we obtain ${A(Z) = 0.0333^{+0.0002}_{-0.0004}}$ (see the red distribution in Figure~\ref{fig:Ahist}). 
This translates into $\delta\sigma_{Z} = ^{+0.6\%}_{-1.2\%}$.

We assume the remaining stellar population-related uncertainties (SL, SPS and IMF) to be independent of $\delta\sigma_{Z,\alpha}$ to first order and, following the approach of \citet{Moresco20}, combine them in quadrature.
This assumption is an approximation, as some of these systematic effects are likely to be correlated.
We obtain a total relative systematic uncertainty of $^{+13\%}_{-11\%}$ (including $[\alpha$/Fe]).

\section{Results and discussion}

\begin{figure*}[ht]
\centering
\includegraphics[width = 0.98\textwidth]{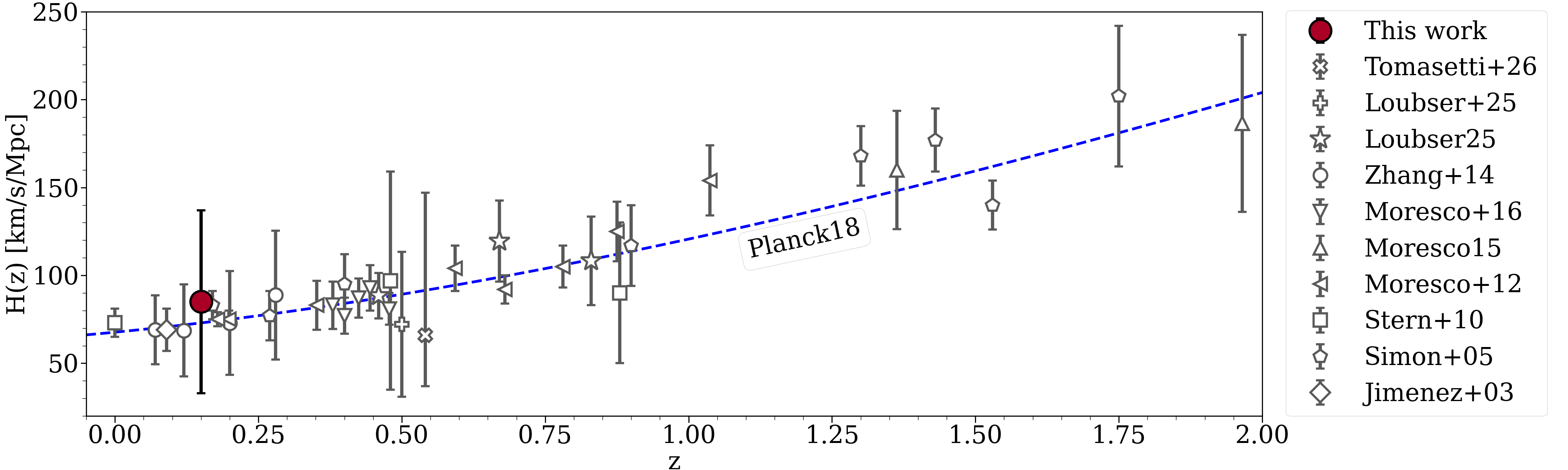}
\caption{$H(z)$ as a function of redshift. Empty symbols show estimates from previous CC studies, as listed at the right side of the plot (\citealp{Jimenez03, Simon05, Stern10, Moresco12,Moresco15, Moresco16, Zhang14, Loubser25, Loubser25+,Tomasetti25}). The blue dashed line presents the flat $\Lambda$CDM trend from \cite{Planck}.}
\label{fig:Hz}
\end{figure*}

Using relic galaxies as CCs for the first time, we measure the Hubble parameter at the median redshift of our sample ($z=0.15$):
\begin{equation}
    \textrm{H}(z=0.15) = 85.0\pm51.5 \textrm{ [stat] }^{+10.2}_{-8.9} \textrm{ [sys] km\,s}^{-1}\,\textrm{Mpc}^{-1}.
\end{equation}

The result is fully consistent, within uncertainties, with previous CC measurements as well as with the prediction of the flat $\Lambda$CDM cosmology inferred from \textit{Planck} \citep{Planck}, as shown in Figure~\ref{fig:Hz}. 
Although the total uncertainty remains relatively large (62\%), this level of precision is comparable to that achieved by recent CC analyses based on similarly sized samples of passive galaxies. 
For example, \cite{Loubser25+} reported a relative uncertainty of $\sim47$\% using 55 brightest cluster galaxies, while \cite{Tomasetti25} obtained uncertainties reaching $\sim124$\% for a sample of 38 cluster ellipticals. 
In this context, our measurement already demonstrates that relic-selected chronometers provide competitive constraints despite the still limited sample size.

The most important result of this work is not the absolute precision of the current $H(z)$ estimate itself, but rather the demonstration that relic galaxies substantially reduce the astrophysical systematic uncertainties affecting the CC method. 
In standard CC analyses, large uncertainties arise from extended or complex SFHs, rejuvenation episodes, merger-driven stellar mixing, and degeneracies in stellar population modelling. 
Relic galaxies are uniquely advantageous because they minimise all these effects simultaneously. 
Their extremely compact morphology, absence of significant late-time accretion, and highly synchronised early assembly histories make them much closer to the idealised population originally envisioned for the CC method.

This is connected to the remarkably small $Z$-related uncertainty that we obtain. 
By propagating the measured $Z$ distribution of the final sample through the SPS calibration, we estimate a contribution of only $\delta\sigma_Z \leq 1.2$\%. 
This value is lower by approximately a factor of 2.5 than the most restrictive estimates currently available in the literature \citep{Loubser25+}, and almost an order of magnitude smaller than the uncertainties typically expected for the general population of massive passive galaxies \citep[$\sim10$\%;][]{Moresco20}. 

The origin of this reduction is physically meaningful. 
The $Z$ of the relic sample occupy a very specific region of the $A(Z)$ relation, namely close to the turnover point shown in the bottom panel of Figure~\ref{fig:analysisA}, where the slope $A \equiv \mbox{d}D_n4000/\mbox{d}t$ becomes nearly insensitive to the $Z$ variations. 
As a consequence, uncertainties in $Z$ propagate only weakly into the calibration of the differential age relation. 
This behaviour appears to be a natural consequence of the uniformly high $Z$ and homogeneous stellar populations characteristic of relic systems.

Assuming the standard framework adopted in most CC studies, namely fixed [$\alpha$/Fe], we estimate a total systematic uncertainty of $\delta\sigma_{\rm sys}\leq8.7$\%.

This value is already among the lowest currently reported in the CC literature, comparing favourably with the $\sim20$\% reported by \citet{Tomasetti25}, $\sim10$\% by \citet{Loubser25+}, $\gtrsim10$\% by \citet{Alvarez25}, 20\% by \citet{Tomasetti23}, $\geq10$\% by \citet{Jiao23}, 50\% by \citet{Ratsimbazafy17}, and $\geq11$\% by \citet[][who report that the statistical uncertainty can be reduced to $\sim6$\% when the full sample is analysed jointly]{Moresco16}.
A key result of this work is that the commonly adopted assumption of fixed or homogeneous [$\alpha$/Fe] can lead to an underestimate of the systematic uncertainty associated with the $D_n4000$ method.
At moderate metallicities ($Z \leq 1.5\,Z_\odot$), the impact of [$\alpha$/Fe] variations is comparable to that of metallicity itself.
However, above this threshold, the sensitivity of the $D_n4000$--age calibration to [$\alpha$/Fe] increases rapidly and becomes the dominant stellar population-related systematic.
When the observed [$\alpha$/Fe] distribution is explicitly propagated through the SPS calibration, the total systematic uncertainty increases from $\sim8.7$\% to $\sim13$\%.

This result has important implications for future CC analyses. 
Current CC pipelines generally do not propagate the explicit contribution of $\alpha$-enhancement to the systematic budget, following the fact that massive quiescent galaxies are well known to exhibit super-solar abundance ratios. 
Our analysis indicates that an additional contribution of approximately $\delta\sigma_\alpha \simeq 9$\% should be included whenever [$\alpha$/Fe] variations are not explicitly modelled. 
This effect is particularly relevant for old and metal-rich stellar populations, precisely the regime where CC measurements are expected to perform best.

At the same time, our analysis also highlights current limitations in SPS modelling.
The posterior distribution of $A(Z,[\alpha/\mathrm{Fe}])$ is mildly asymmetric (Appendix~\ref{app:Adistr}), suggesting that the inferred uncertainty may itself still be slightly underestimated.
This asymmetry most likely reflects the limited coverage of extremely $\alpha$-enhanced stellar populations in current SPS libraries, which typically do not extend beyond [$\alpha$/Fe] $\gtrsim 0.4$. 
To avoid extrapolating beyond the available SPS models, Monte Carlo realizations outside this range are discarded when estimating the uncertainty in $A$, which may lead to a slight underestimate of the final systematic uncertainty.
Improved SPS models with both a wider and denser sampling of chemical abundance patterns will therefore be essential to fully exploit the potential of relic-based chronometers.

Despite the significant reduction in systematics, the total uncertainty of the present measurement remains dominated by the statistical component. 
This is expected given the relatively small size of the current relic sample. 
The primary limitation of the method is therefore no longer the astrophysical purity of the tracers, but rather the number of available objects. 
This is a particularly encouraging result because statistical uncertainties can be straightforwardly reduced with future surveys.


Our measurement should therefore be interpreted primarily as a proof-of-concept demonstrating that relic galaxies constitute one of the most promising populations currently available for the CC method. 
The present analysis establishes both the feasibility of the approach and the systematic advantages of relic-selected chronometers, opening the way towards a new generation of CC measurements based on physically cleaner tracers.

\section{Summary}
In this work, we presented the first measurement of the Hubble parameter obtained using relic galaxies as cosmic chronometers (CCs), based on carefully selected sample of extremely passive ultra-compact massive galaxies (UCMGs). 
Although the current uncertainty remains relatively large, the result is fully consistent with previous CC measurements and with the expectations from the standard $\Lambda$CDM cosmology. 
More importantly, this work demonstrates that relic galaxies provide a particularly powerful population for the CC method because they naturally minimise several of the dominant astrophysical systematic effects affecting standard passive galaxy samples.

The main strength of relic galaxies lies in their exceptionally simple evolutionary histories. 
Their compact morphology and absence of significant merger activity imply that they experienced extremely rapid and synchronised star formation at early cosmic times, followed by nearly passive evolution over most of the age of the Universe. 
This substantially reduces the impact of extended or rejuvenated SFHs, which are important sources of uncertainty in traditional CC analyses. 
At the same time, the uniformly high $Z$ of relics place them close to the turnover region of the $A(Z)$ relation, where the $D_n4000$--age calibration becomes weakly sensitive to $Z$ variations. 
As a consequence, the $Z$-related contribution to the systematic uncertainty budget is strongly suppressed.
We therefore obtain a total relative systematic uncertainty of $\delta\sigma_{\rm sys}\leq12.5$\%. 
If variations in [$\alpha$/Fe] are neglected, consistently with the assumptions commonly adopted in the literature, the systematic uncertainty decreases to $\delta\sigma_{\rm sys}\leq8.7$\%, among the lowest values currently achieved in CC studies.


An important result of this work is the identification of $\alpha$-enhancement as a major contributor to the systematic uncertainty budget. 
While previous CC analyses have primarily focused on the role of $Z$, we show that [$\alpha$/Fe] variations become increasingly important in the metal-rich regime characteristic of relic systems, eventually dominating the uncertainty on the $D_n4000$--age calibration. 
This indicates that future high-precision CC studies should explicitly account for abundance-ratio variations and motivates the development of SPS models with a wider and denser sampling of chemical enrichment histories.


At present, the dominant limitation of the method is statistical rather than systematic, driven by the still limited number of confirmed relic galaxies with high-quality spectroscopy. 
However, this situation is expected to improve rapidly over the next decade. 
Wide-field imaging surveys such as \textit{Euclid} and the Vera C. Rubin Observatory will dramatically increase the number of UCMG candidates, while spectroscopic facilities such as DESI and 4MOST will provide the data required for robust relic confirmation and stellar population analysis. 
These datasets will allow relic-based CC analyses to move from the current proof-of-concept stage towards competitive precision cosmology.
Overall, our results demonstrate that relic galaxies represent an exceptionally promising population for the CC method. 
By combining highly homogeneous stellar populations, minimal evolutionary contamination, and reduced systematic uncertainties, relics provide a physically cleaner route towards direct measurements of the expansion history of the Universe.

\section{Data availability}
The data underlying this article will be shared on request.

\begin{acknowledgements} 
We would like to thank the anonymous referee for the careful reading of the manuscript and the very useful comments.
K.L. acknowledges support of the Polish Ministry of Education and Science through the grant PN/01/0034/2022 under ‘Perły Nauki’ programme. 
A.P. and M.B. were supported by the Polish National Science Centre grant 2023/50/A/ST9/00579.
D.D. acknowledges support from the Polish National Agency for Academic Exchange (Bekker grant BPN /BEK/2024/1/00029/DEC/1).
P.M. acknowledges the support of the Polish National Science Center project UMO-2024/53/B/ST9/00230.
We acknowledge the support of the CosmoVerse COST action.
\end{acknowledgements}

\bibliographystyle{aa}
\bibliography{main}

@ARTICLE{DeLucia06,
       author = {{De Lucia}, Gabriella and {Springel}, Volker and {White}, Simon D.~M. and {Croton}, Darren and {Kauffmann}, Guinevere},
        title = "{The formation history of elliptical galaxies}",
      journal = {\mnras},
         year = 2006,
        month = feb,
       volume = {366},
       number = {2},
        pages = {499-509},
          doi = {10.1111/j.1365-2966.2005.09879.x},
archivePrefix = {arXiv},
       eprint = {astro-ph/0509725},
 primaryClass = {astro-ph},
       adsurl = {https://ui.adsabs.harvard.edu/abs/2006MNRAS.366..499D}
}

@INPROCEEDINGS{Tomre77,
       author = {{Toomre}, Alar},
        title = "{Mergers and Some Consequences}",
    booktitle = {Evolution of Galaxies and Stellar Populations},
         year = 1977,
       editor = {{Tinsley}, Beatrice M. and {Larson}, D. Campbell, Richard B. Gehret},
        month = jan,
        pages = {401},
       adsurl = {https://ui.adsabs.harvard.edu/abs/1977egsp.conf..401T}
}

@ARTICLE{Damjanov+11,
   author = {{Damjanov}, I. and {Abraham}, R.~G. and {Glazebrook}, K. and 
	{McCarthy}, P.~J. and {Caris}, E. and {Carlberg}, R.~G. and 
	{Chen}, H.-W. and {Crampton}, D. and {Green}, A.~W. and {J{\o}rgensen}, I. and 
	{Juneau}, S. and {Le Borgne}, D. and {Marzke}, R.~O. and {Mentuch}, E. and 
	{Murowinski}, R. and {Roth}, K. and {Savaglio}, S. and {Yan}, H.
	},
    title = "{Red Nuggets at High Redshift: Structural Evolution of Quiescent Galaxies Over 10 Gyr of Cosmic History}",
  journal = {\apjl},
archivePrefix = "arXiv",
   eprint = {1108.0656},
     year = 2011,
    month = oct,
   volume = 739,
      eid = {L44},
    pages = {L44},
      doi = {10.1088/2041-8205/739/2/L44},
   adsurl = {http://adsabs.harvard.edu/abs/2011ApJ...739L..44D}
}

@ARTICLE{Poggianti97,
       author = {{Poggianti}, B.~M. and {Barbaro}, G.},
        title = "{Indicators of star formation: 4000 {\r{A}} break and Balmer lines.}",
      journal = {\aap},
         year = 1997,
        month = sep,
       volume = {325},
        pages = {1025-1030},
          doi = {10.48550/arXiv.astro-ph/9703067},
archivePrefix = {arXiv},
       eprint = {astro-ph/9703067},
 primaryClass = {astro-ph},
       adsurl = {https://ui.adsabs.harvard.edu/abs/1997A&A...325.1025P}
}

@ARTICLE{planck16,
       author = {{Planck Collaboration} and {Ade}, P.~A.~R. and {Aghanim}, N. and {Arnaud}, M. and {Ashdown}, M. and {Aumont}, J. and {Baccigalupi}, C. and {Banday}, A.~J. and {Barreiro}, R.~B. and {Bartlett}, J.~G. and {Bartolo}, N. and {Battaner}, E. and {Battye}, R. and {Benabed}, K. and {Beno{\^\i}t}, A. and {Benoit-L{\'e}vy}, A. and {Bernard}, J.-P. and {Bersanelli}, M. and {Bielewicz}, P. and {Bock}, J.~J. and {Bonaldi}, A. and {Bonavera}, L. and {Bond}, J.~R. and {Borrill}, J. and {Bouchet}, F.~R. and {Boulanger}, F. and {Bucher}, M. and {Burigana}, C. and {Butler}, R.~C. and {Calabrese}, E. and {Cardoso}, J.-F. and {Catalano}, A. and {Challinor}, A. and {Chamballu}, A. and {Chary}, R.-R. and {Chiang}, H.~C. and {Chluba}, J. and {Christensen}, P.~R. and {Church}, S. and {Clements}, D.~L. and {Colombi}, S. and {Colombo}, L.~P.~L. and {Combet}, C. and {Coulais}, A. and {Crill}, B.~P. and {Curto}, A. and {Cuttaia}, F. and {Danese}, L. and {Davies}, R.~D. and {Davis}, R.~J. and {de Bernardis}, P. and {de Rosa}, A. and {de Zotti}, G. and {Delabrouille}, J. and {D{\'e}sert}, F.-X. and {Di Valentino}, E. and {Dickinson}, C. and {Diego}, J.~M. and {Dolag}, K. and {Dole}, H. and {Donzelli}, S. and {Dor{\'e}}, O. and {Douspis}, M. and {Ducout}, A. and {Dunkley}, J. and {Dupac}, X. and {Efstathiou}, G. and {Elsner}, F. and {En{\ss}lin}, T.~A. and {Eriksen}, H.~K. and {Farhang}, M. and {Fergusson}, J. and {Finelli}, F. and {Forni}, O. and {Frailis}, M. and {Fraisse}, A.~A. and {Franceschi}, E. and {Frejsel}, A. and {Galeotta}, S. and {Galli}, S. and {Ganga}, K. and {Gauthier}, C. and {Gerbino}, M. and {Ghosh}, T. and {Giard}, M. and {Giraud-H{\'e}raud}, Y. and {Giusarma}, E. and {Gjerl{\o}w}, E. and {Gonz{\'a}lez-Nuevo}, J. and {G{\'o}rski}, K.~M. and {Gratton}, S. and {Gregorio}, A. and {Gruppuso}, A. and {Gudmundsson}, J.~E. and {Hamann}, J. and {Hansen}, F.~K. and {Hanson}, D. and {Harrison}, D.~L. and {Helou}, G. and {Henrot-Versill{\'e}}, S. and {Hern{\'a}ndez-Monteagudo}, C. and {Herranz}, D. and {Hildebrandt}, S.~R. and {Hivon}, E. and {Hobson}, M. and {Holmes}, W.~A. and {Hornstrup}, A. and {Hovest}, W. and {Huang}, Z. and {Huffenberger}, K.~M. and {Hurier}, G. and {Jaffe}, A.~H. and {Jaffe}, T.~R. and {Jones}, W.~C. and {Juvela}, M. and {Keih{\"a}nen}, E. and {Keskitalo}, R. and {Kisner}, T.~S. and {Kneissl}, R. and {Knoche}, J. and {Knox}, L. and {Kunz}, M. and {Kurki-Suonio}, H. and {Lagache}, G. and {L{\"a}hteenm{\"a}ki}, A. and {Lamarre}, J.-M. and {Lasenby}, A. and {Lattanzi}, M. and {Lawrence}, C.~R. and {Leahy}, J.~P. and {Leonardi}, R. and {Lesgourgues}, J. and {Levrier}, F. and {Lewis}, A. and {Liguori}, M. and {Lilje}, P.~B. and {Linden-V{\o}rnle}, M. and {L{\'o}pez-Caniego}, M. and {Lubin}, P.~M. and {Mac{\'\i}as-P{\'e}rez}, J.~F. and {Maggio}, G. and {Maino}, D. and {Mandolesi}, N. and {Mangilli}, A. and {Marchini}, A. and {Maris}, M. and {Martin}, P.~G. and {Martinelli}, M. and {Mart{\'\i}nez-Gonz{\'a}lez}, E. and {Masi}, S. and {Matarrese}, S. and {McGehee}, P. and {Meinhold}, P.~R. and {Melchiorri}, A. and {Melin}, J.-B. and {Mendes}, L. and {Mennella}, A. and {Migliaccio}, M. and {Millea}, M. and {Mitra}, S. and {Miville-Desch{\^e}nes}, M.-A. and {Moneti}, A. and {Montier}, L. and {Morgante}, G. and {Mortlock}, D. and {Moss}, A. and {Munshi}, D. and {Murphy}, J.~A. and {Naselsky}, P. and {Nati}, F. and {Natoli}, P. and {Netterfield}, C.~B. and {N{\o}rgaard-Nielsen}, H.~U. and {Noviello}, F. and {Novikov}, D. and {Novikov}, I. and {Oxborrow}, C.~A. and {Paci}, F. and {Pagano}, L. and {Pajot}, F. and {Paladini}, R. and {Paoletti}, D. and {Partridge}, B. and {Pasian}, F. and {Patanchon}, G. and {Pearson}, T.~J. and {Perdereau}, O. and {Perotto}, L. and {Perrotta}, F. and {Pettorino}, V. and {Piacentini}, F. and {Piat}, M. and {Pierpaoli}, E. and {Pietrobon}, D. and {Plaszczynski}, S. and {Pointecouteau}, E. and {Polenta}, G. and {Popa}, L. and {Pratt}, G.~W. and {Pr{\'e}zeau}, G.},
        title = "{Planck 2015 results. XIII. Cosmological parameters}",
      journal = {\aap},
         year = 2016,
        month = sep,
       volume = {594},
          eid = {A13},
        pages = {A13},
          doi = {10.1051/0004-6361/201525830},
archivePrefix = {arXiv},
       eprint = {1502.01589},
 primaryClass = {astro-ph.CO},
       adsurl = {https://ui.adsabs.harvard.edu/abs/2016A&A...594A..13P}
}

@ARTICLE{Comeron+23,
       author = {{Comer{\'o}n}, S{\'e}bastien and {Trujillo}, Ignacio and {Cappellari}, Michele and {Buitrago}, Fernando and {Gardu{\~n}o}, Luis E. and {Zaragoza-Cardiel}, Javier and {Zinchenko}, Igor A. and {Lara-L{\'o}pez}, Maritza A. and {Ferr{\'e}-Mateu}, Anna and {Dib}, Sami},
        title = "{The massive relic galaxy NGC 1277 is dark matter deficient. From dynamical models of integral-field stellar kinematics out to five effective radii}",
      journal = {\aap},
         year = 2023,
        month = jul,
       volume = {675},
          eid = {A143},
        pages = {A143},
          doi = {10.1051/0004-6361/202346291},
archivePrefix = {arXiv},
       eprint = {2303.11360},
 primaryClass = {astro-ph.GA},
       adsurl = {https://ui.adsabs.harvard.edu/abs/2023A&A...675A.143C}
}

@ARTICLE{Pascalau26,
       author = {{Pascalau}, Robert G. and {D'Eugenio}, Francesco and {Tacchella}, Sandro and {Maiolino}, Roberto and {Cappellari}, Michele and {Duan}, Qiao and {Lagos}, Claudia del P. and {Bunker}, Andrew J. and {Jones}, Gareth C. and {Scholtz}, Jan and {{\"U}bler}, Hannah and {Cresci}, Giovanni and {Arribas}, Santiago and {Perna}, Michele and {van der Wel}, Arjen and {Danhaive}, A. Lola and {McClymont}, William and {Williams}, Christina C. and {de Graaff}, Anna and {Vani}, Akash and {Maseda}, Michael V. and {Carnall}, Adam C. and {Charlot}, St{\'e}phane and {Carniani}, Stefano and {Goh}, Tze P. and {Ji}, Zhiyuan and {P{\'e}rez Gonz{\'a}lez}, Pablo},
        title = "{When relics were made: vigorous stellar rotation and low dark matter content in the massive ultra-compact galaxy GS-9209 at z = 4.66}",
      journal = {\mnras},
         year = 2026,
        month = mar,
       volume = {547},
       number = {1},
          eid = {stag210},
        pages = {stag210},
          doi = {10.1093/mnras/stag210},
archivePrefix = {arXiv},
       eprint = {2505.06349},
 primaryClass = {astro-ph.GA},
       adsurl = {https://ui.adsabs.harvard.edu/abs/2026MNRAS.547ag210P}
}

@ARTICLE{Mills25,
       author = {{Mills}, John and {Spiniello}, Chiara and {Sergeyev}, Alexey and {Tortora}, Crescenzo and {Khramtsov}, Vladyslav and {D'Ago}, Giuseppe and {Maksymowicz-Maciata}, Michalina and {Benedetti}, Jo{\~a}o P.~V. and {Ferr{\'e}-Mateu}, Anna and {Cappellari}, Michele and {Davies}, Roger and {Hartke}, Johanna and {Rosen}, Charles},
        title = "{E-INSPIRE ─ I. Bridging the gap with the local Universe: stellar population of a statistical sample of ultra-compact massive galaxies at z < 0.3}",
      journal = {\mnras},
         year = 2025,
        month = aug,
       volume = {541},
       number = {3},
        pages = {2440-2458},
          doi = {10.1093/mnras/staf516},
archivePrefix = {arXiv},
       eprint = {2501.16126},
 primaryClass = {astro-ph.GA},
       adsurl = {https://ui.adsabs.harvard.edu/abs/2025MNRAS.541.2440M}
}

@ARTICLE{ferre-mateu17,
       author = {{Ferr{\'e}-Mateu}, Anna and {Trujillo}, Ignacio and {Mart{\'\i}n-Navarro}, Ignacio and {Vazdekis}, Alexandre and {Mezcua}, Mar and {Balcells}, Marc and {Dom{\'\i}nguez}, Lilian},
        title = "{Two new confirmed massive relic galaxies: red nuggets in the present-day Universe}",
      journal = {\mnras},
         year = 2017,
        month = may,
       volume = {467},
       number = {2},
        pages = {1929-1939},
          doi = {10.1093/mnras/stx171},
archivePrefix = {arXiv},
       eprint = {1701.05197},
 primaryClass = {astro-ph.GA},
       adsurl = {https://ui.adsabs.harvard.edu/abs/2017MNRAS.467.1929F}
}

@ARTICLE{Cappellari04,
       author = {{Cappellari}, Michele and {Emsellem}, Eric},
        title = "{Parametric Recovery of Line-of-Sight Velocity Distributions from Absorption-Line Spectra of Galaxies via Penalized Likelihood}",
      journal = {\pasp},
         year = 2004,
        month = feb,
       volume = {116},
       number = {816},
        pages = {138-147},
          doi = {10.1086/381875},
archivePrefix = {arXiv},
       eprint = {astro-ph/0312201},
 primaryClass = {astro-ph},
       adsurl = {https://ui.adsabs.harvard.edu/abs/2004PASP..116..138C}
}

@ARTICLE{Cappellari17,
       author = {{Cappellari}, Michele},
        title = "{Improving the full spectrum fitting method: accurate convolution with Gauss-Hermite functions}",
      journal = {\mnras},
         year = 2017,
        month = apr,
       volume = {466},
       number = {1},
        pages = {798-811},
          doi = {10.1093/mnras/stw3020},
archivePrefix = {arXiv},
       eprint = {1607.08538},
 primaryClass = {astro-ph.GA},
       adsurl = {https://ui.adsabs.harvard.edu/abs/2017MNRAS.466..798C}
}

@ARTICLE{Cappellari23,
       author = {{Cappellari}, Michele},
        title = "{Full spectrum fitting with photometry in PPXF: stellar population versus dynamical masses, non-parametric star formation history and metallicity for 3200 LEGA-C galaxies at redshift z {\ensuremath{\approx}} 0.8}",
      journal = {\mnras},
         year = 2023,
        month = dec,
       volume = {526},
       number = {3},
        pages = {3273-3300},
          doi = {10.1093/mnras/stad2597},
archivePrefix = {arXiv},
       eprint = {2208.14974},
 primaryClass = {astro-ph.GA},
       adsurl = {https://ui.adsabs.harvard.edu/abs/2023MNRAS.526.3273C}
}

@ARTICLE{Vazdekis16,
       author = {{Vazdekis}, A. and {Koleva}, M. and {Ricciardelli}, E. and {R{\"o}ck}, B. and {Falc{\'o}n-Barroso}, J.},
        title = "{UV-extended E-MILES stellar population models: young components in massive early-type galaxies}",
      journal = {\mnras},
         year = 2016,
        month = dec,
       volume = {463},
       number = {4},
        pages = {3409-3436},
          doi = {10.1093/mnras/stw2231},
archivePrefix = {arXiv},
       eprint = {1612.01187},
 primaryClass = {astro-ph.GA},
       adsurl = {https://ui.adsabs.harvard.edu/abs/2016MNRAS.463.3409V}
}

@ARTICLE{Moresco12,
       author = {{Moresco}, M. and {Cimatti}, A. and {Jimenez}, R. and {Pozzetti}, L. and {Zamorani}, G. and {Bolzonella}, M. and {Dunlop}, J. and {Lamareille}, F. and {Mignoli}, M. and {Pearce}, H. and {Rosati}, P. and {Stern}, D. and {Verde}, L. and {Zucca}, E. and {Carollo}, C.~M. and {Contini}, T. and {Kneib}, J.-P. and {Le F{\`e}vre}, O. and {Lilly}, S.~J. and {Mainieri}, V. and {Renzini}, A. and {Scodeggio}, M. and {Balestra}, I. and {Gobat}, R. and {McLure}, R. and {Bardelli}, S. and {Bongiorno}, A. and {Caputi}, K. and {Cucciati}, O. and {de la Torre}, S. and {de Ravel}, L. and {Franzetti}, P. and {Garilli}, B. and {Iovino}, A. and {Kampczyk}, P. and {Knobel}, C. and {Kova{\v{c}}}, K. and {Le Borgne}, J.-F. and {Le Brun}, V. and {Maier}, C. and {Pell{\'o}}, R. and {Peng}, Y. and {Perez-Montero}, E. and {Presotto}, V. and {Silverman}, J.~D. and {Tanaka}, M. and {Tasca}, L.~A.~M. and {Tresse}, L. and {Vergani}, D. and {Almaini}, O. and {Barnes}, L. and {Bordoloi}, R. and {Bradshaw}, E. and {Cappi}, A. and {Chuter}, R. and {Cirasuolo}, M. and {Coppa}, G. and {Diener}, C. and {Foucaud}, S. and {Hartley}, W. and {Kamionkowski}, M. and {Koekemoer}, A.~M. and {L{\'o}pez-Sanjuan}, C. and {McCracken}, H.~J. and {Nair}, P. and {Oesch}, P. and {Stanford}, A. and {Welikala}, N.},
        title = "{Improved constraints on the expansion rate of the Universe up to z \raisebox{-0.5ex}\textasciitilde 1.1 from the spectroscopic evolution of cosmic chronometers}",
      journal = {\jcap},
         year = 2012,
        month = aug,
       volume = {2012},
       number = {8},
          eid = {006},
        pages = {006},
          doi = {10.1088/1475-7516/2012/08/006},
archivePrefix = {arXiv},
       eprint = {1201.3609},
 primaryClass = {astro-ph.CO},
       adsurl = {https://ui.adsabs.harvard.edu/abs/2012JCAP...08..006M}
}

@ARTICLE{Moresco15,
       author = {{Moresco}, M.},
        title = "{Raising the bar: new constraints on the Hubble parameter with cosmic chronometers at z \raisebox{-0.5ex}\textasciitilde 2.}",
      journal = {\mnras},
         year = 2015,
        month = jun,
       volume = {450},
        pages = {L16-L20},
          doi = {10.1093/mnrasl/slv037},
archivePrefix = {arXiv},
       eprint = {1503.01116},
 primaryClass = {astro-ph.CO},
       adsurl = {https://ui.adsabs.harvard.edu/abs/2015MNRAS.450L..16M}
}

@ARTICLE{Loubser25+,
       author = {{Loubser}, S. Ilani and {Alabi}, Adebusola B. and {Hilton}, Matt and {Ma}, Yin-Zhe and {Tang}, Xin and {Hatamkhani}, Narges and {Cress}, Catherine and {Skelton}, Rosalind E. and {Nkosi}, S. Andile},
        title = "{An independent estimate of H(z) at z = 0.5 from the stellar ages of brightest cluster galaxies}",
      journal = {\mnras},
         year = 2025,
        month = jul,
       volume = {540},
       number = {4},
        pages = {3135-3149},
          doi = {10.1093/mnras/staf915},
archivePrefix = {arXiv},
       eprint = {2506.03836},
 primaryClass = {astro-ph.CO},
       adsurl = {https://ui.adsabs.harvard.edu/abs/2025MNRAS.540.3135L}
}

@ARTICLE{Szpila25,
       author = {{Szpila}, Jakub and {Dav{\'e}}, Romeel and {Rennehan}, Douglas and {Cui}, Weiguang and {Hough}, Renier T.},
        title = "{The nature and evolution of early massive quenched galaxies in the SIMBA-C simulation}",
      journal = {\mnras},
         year = 2025,
        month = feb,
       volume = {537},
       number = {2},
        pages = {1849-1868},
          doi = {10.1093/mnras/staf132},
archivePrefix = {arXiv},
       eprint = {2402.08729},
 primaryClass = {astro-ph.GA},
       adsurl = {https://ui.adsabs.harvard.edu/abs/2025MNRAS.537.1849S}
}

@ARTICLE{Chauke19,
       author = {{Chauke}, Priscilla and {van der Wel}, Arjen and {Pacifici}, Camilla and {Bezanson}, Rachel and {Wu}, Po-Feng and {Gallazzi}, Anna and {Straatman}, Caroline and {Franx}, Marijn and {Bari{\v{s}}i{\'c}}, Ivana and {Bell}, Eric F. and {van Houdt}, Josha and {Maseda}, Michael V. and {Muzzin}, Adam and {Sobral}, David and {Spilker}, Justin},
        title = "{Rejuvenation in z {\ensuremath{\sim}} 0.8 Quiescent Galaxies in LEGA-C}",
      journal = {\apj},
         year = 2019,
        month = may,
       volume = {877},
       number = {1},
          eid = {48},
        pages = {48},
          doi = {10.3847/1538-4357/ab164d},
archivePrefix = {arXiv},
       eprint = {1905.07541},
 primaryClass = {astro-ph.GA},
       adsurl = {https://ui.adsabs.harvard.edu/abs/2019ApJ...877...48C}
}

@article{Rosen+26,
  author = {{Rosen}, Charles and {Spiniello}, Chiara and {Mills}, John and {Sergeyev}, Alexey and {Khramtsov}, Vladyslav and {Ferr{\'e}-Mateu}, Anna and {Hartke}, Johanna and {Maskymowicz-Maciata}, Michalina and {Siudek}, Malgorzata and {Tortora}, Crescenzo},
  title = {E-INSPIRE - II. Finding relics from wide-sky multi-band surveys:
A proof-of-concept machine learning regression algorithm},
  year = {2026},
  journal = {submitted to MNRAS},
}

@ARTICLE{Barro13,
       author = {{Barro}, Guillermo and {Faber}, S.~M. and {P{\'e}rez-Gonz{\'a}lez}, Pablo G. and {Koo}, David C. and {Williams}, Christina C. and {Kocevski}, Dale D. and {Trump}, Jonathan R. and {Mozena}, Mark and {McGrath}, Elizabeth and {van der Wel}, Arjen and {Wuyts}, Stijn and {Bell}, Eric F. and {Croton}, Darren J. and {Ceverino}, Daniel and {Dekel}, Avishai and {Ashby}, M.~L.~N. and {Cheung}, Edmond and {Ferguson}, Henry C. and {Fontana}, Adriano and {Fang}, Jerome and {Giavalisco}, Mauro and {Grogin}, Norman A. and {Guo}, Yicheng and {Hathi}, Nimish P. and {Hopkins}, Philip F. and {Huang}, Kuang-Han and {Koekemoer}, Anton M. and {Kartaltepe}, Jeyhan S. and {Lee}, Kyoung-Soo and {Newman}, Jeffrey A. and {Porter}, Lauren A. and {Primack}, Joel R. and {Ryan}, Russell E. and {Rosario}, David and {Somerville}, Rachel S. and {Salvato}, Mara and {Hsu}, Li-Ting},
        title = "{CANDELS: The Progenitors of Compact Quiescent Galaxies at z \raisebox{-0.5ex}\textasciitilde 2}",
      journal = {\apj},
         year = 2013,
        month = mar,
       volume = {765},
       number = {2},
          eid = {104},
        pages = {104},
          doi = {10.1088/0004-637X/765/2/104},
archivePrefix = {arXiv},
       eprint = {1206.5000},
 primaryClass = {astro-ph.CO},
       adsurl = {https://ui.adsabs.harvard.edu/abs/2013ApJ...765..104B}
}

@ARTICLE{Perivolaropoulos22,
       author = {{Perivolaropoulos}, L. and {Skara}, F.},
        title = "{Challenges for {\ensuremath{\Lambda}}CDM: An update}",
      journal = {\nar},
         year = 2022,
        month = dec,
       volume = {95},
          eid = {101659},
        pages = {101659},
          doi = {10.1016/j.newar.2022.101659},
archivePrefix = {arXiv},
       eprint = {2105.05208},
 primaryClass = {astro-ph.CO},
       adsurl = {https://ui.adsabs.harvard.edu/abs/2022NewAR..9501659P}
}

@ARTICLE{Shah21,
       author = {{Shah}, Paul and {Lemos}, Pablo and {Lahav}, Ofer},
        title = "{A buyer's guide to the Hubble constant}",
      journal = {\aapr},
         year = 2021,
        month = dec,
       volume = {29},
       number = {1},
          eid = {9},
        pages = {9},
          doi = {10.1007/s00159-021-00137-4},
archivePrefix = {arXiv},
       eprint = {2109.01161},
 primaryClass = {astro-ph.CO},
       adsurl = {https://ui.adsabs.harvard.edu/abs/2021A&ARv..29....9S}
}

@ARTICLE{Bernal16,
       author = {{Bernal}, Jos{\'e} Luis and {Verde}, Licia and {Riess}, Adam G.},
        title = "{The trouble with H$_{0}$}",
      journal = {\jcap},
         year = 2016,
        month = oct,
       volume = {2016},
       number = {10},
          eid = {019},
        pages = {019},
          doi = {10.1088/1475-7516/2016/10/019},
archivePrefix = {arXiv},
       eprint = {1607.05617},
 primaryClass = {astro-ph.CO},
       adsurl = {https://ui.adsabs.harvard.edu/abs/2016JCAP...10..019B}
}

@ARTICLE{Hinshaw13,
       author = {{Hinshaw}, G. and {Larson}, D. and {Komatsu}, E. and {Spergel}, D.~N. and {Bennett}, C.~L. and {Dunkley}, J. and {Nolta}, M.~R. and {Halpern}, M. and {Hill}, R.~S. and {Odegard}, N. and {Page}, L. and {Smith}, K.~M. and {Weiland}, J.~L. and {Gold}, B. and {Jarosik}, N. and {Kogut}, A. and {Limon}, M. and {Meyer}, S.~S. and {Tucker}, G.~S. and {Wollack}, E. and {Wright}, E.~L.},
        title = "{Nine-year Wilkinson Microwave Anisotropy Probe (WMAP) Observations: Cosmological Parameter Results}",
      journal = {\apjs},
         year = 2013,
        month = oct,
       volume = {208},
       number = {2},
          eid = {19},
        pages = {19},
          doi = {10.1088/0067-0049/208/2/19},
archivePrefix = {arXiv},
       eprint = {1212.5226},
 primaryClass = {astro-ph.CO},
       adsurl = {https://ui.adsabs.harvard.edu/abs/2013ApJS..208...19H}
}

@ARTICLE{Borghi22,
       author = {{Borghi}, Nicola and {Moresco}, Michele and {Cimatti}, Andrea},
        title = "{Toward a Better Understanding of Cosmic Chronometers: A New Measurement of H(z) at z   0.7}",
      journal = {\apjl},
         year = 2022,
        month = mar,
       volume = {928},
       number = {1},
          eid = {L4},
        pages = {L4},
          doi = {10.3847/2041-8213/ac3fb2},
archivePrefix = {arXiv},
       eprint = {2110.04304},
 primaryClass = {astro-ph.CO},
       adsurl = {https://ui.adsabs.harvard.edu/abs/2022ApJ...928L...4B}
}

@ARTICLE{Jiao23,
       author = {{Jiao}, Kang and {Borghi}, Nicola and {Moresco}, Michele and {Zhang}, Tong-Jie},
        title = "{New Observational H(z) Data from Full-spectrum Fitting of Cosmic Chronometers in the LEGA-C Survey}",
      journal = {\apjs},
         year = 2023,
        month = apr,
       volume = {265},
       number = {2},
          eid = {48},
        pages = {48},
          doi = {10.3847/1538-4365/acbc77},
archivePrefix = {arXiv},
       eprint = {2205.05701},
 primaryClass = {astro-ph.CO},
       adsurl = {https://ui.adsabs.harvard.edu/abs/2023ApJS..265...48J}
}

@ARTICLE{Ratsimbazafy17,
       author = {{Ratsimbazafy}, A.~L. and {Loubser}, S.~I. and {Crawford}, S.~M. and {Cress}, C.~M. and {Bassett}, B.~A. and {Nichol}, R.~C. and {V{\"a}is{\"a}nen}, P.},
        title = "{Age-dating luminous red galaxies observed with the Southern African Large Telescope}",
      journal = {\mnras},
         year = 2017,
        month = may,
       volume = {467},
       number = {3},
        pages = {3239-3254},
          doi = {10.1093/mnras/stx301},
archivePrefix = {arXiv},
       eprint = {1702.00418},
 primaryClass = {astro-ph.CO},
       adsurl = {https://ui.adsabs.harvard.edu/abs/2017MNRAS.467.3239R}
}

@article{ Moresco18,
author = { Michele Moresco and Raul Jimenez and Licia Verde and Lucia Pozzetti and Andrea Cimatti and Annalisa Citro },
title = { Setting the Stage for Cosmic Chronometers. I. Assessing the Impact of Young Stellar Populations on Hubble Parameter Measurements },
eprint = { Arxiv:1804.05864v2 },
journal = { The Astrophysical Journal },
volume = { 868 },
pages = { 84 },
year = { 2018 },
month = { 11 },
url = { https://arxiv.org/abs/1804.05864 },
}

@ARTICLE{Johnson21,
       author = {{Johnson}, Benjamin D. and {Leja}, Joel and {Conroy}, Charlie and {Speagle}, Joshua S.},
        title = "{Stellar Population Inference with Prospector}",
      journal = {\apjs},
         year = 2021,
        month = jun,
       volume = {254},
       number = {2},
          eid = {22},
        pages = {22},
          doi = {10.3847/1538-4365/abef67},
archivePrefix = {arXiv},
       eprint = {2012.01426},
 primaryClass = {astro-ph.GA},
       adsurl = {https://ui.adsabs.harvard.edu/abs/2021ApJS..254...22J}
}

@ARTICLE{Carnall18,
       author = {{Carnall}, A.~C. and {McLure}, R.~J. and {Dunlop}, J.~S. and {Dav{\'e}}, R.},
        title = "{Inferring the star formation histories of massive quiescent galaxies with BAGPIPES: evidence for multiple quenching mechanisms}",
      journal = {\mnras},
         year = 2018,
        month = nov,
       volume = {480},
       number = {4},
        pages = {4379-4401},
          doi = {10.1093/mnras/sty2169},
archivePrefix = {arXiv},
       eprint = {1712.04452},
 primaryClass = {astro-ph.GA},
       adsurl = {https://ui.adsabs.harvard.edu/abs/2018MNRAS.480.4379C}
}

@ARTICLE{Moresco20,
       author = {{Moresco}, Michele and {Jimenez}, Raul and {Verde}, Licia and {Cimatti}, Andrea and {Pozzetti}, Lucia},
        title = "{Setting the Stage for Cosmic Chronometers. II. Impact of Stellar Population Synthesis Models Systematics and Full Covariance Matrix}",
      journal = {\apj},
         year = 2020,
        month = jul,
       volume = {898},
       number = {1},
          eid = {82},
        pages = {82},
          doi = {10.3847/1538-4357/ab9eb0},
archivePrefix = {arXiv},
       eprint = {2003.07362},
 primaryClass = {astro-ph.GA},
       adsurl = {https://ui.adsabs.harvard.edu/abs/2020ApJ...898...82M}
}

@ARTICLE{Kaushal24,
       author = {{Kaushal}, Yasha and {Nersesian}, Angelos and {Bezanson}, Rachel and {van der Wel}, Arjen and {Leja}, Joel and {Carnall}, Adam and {Gallazzi}, Anna and {Zibetti}, Stefano and {Khullar}, Gourav and {Franx}, Marijn and {Muzzin}, Adam and {de Graaff}, Anna and {Pacifici}, Camilla and {Whitaker}, Katherine E. and {Bell}, Eric F. and {Martorano}, Marco},
        title = "{A Census of Star Formation Histories of Massive Galaxies at 0.6 < z < 1 from Spectrophotometric Modeling Using Bagpipes and Prospector}",
      journal = {\apj},
         year = 2024,
        month = jan,
       volume = {961},
       number = {1},
          eid = {118},
        pages = {118},
          doi = {10.3847/1538-4357/ad0c4e},
archivePrefix = {arXiv},
       eprint = {2307.03725},
 primaryClass = {astro-ph.GA},
       adsurl = {https://ui.adsabs.harvard.edu/abs/2024ApJ...961..118K}
}

@ARTICLE{dave19,
       author = {{Dav{\'e}}, Romeel and {Angl{\'e}s-Alc{\'a}zar}, Daniel and {Narayanan}, Desika and {Li}, Qi and {Rafieferantsoa}, Mika H. and {Appleby}, Sarah},
        title = "{SIMBA: Cosmological simulations with black hole growth and feedback}",
      journal = {\mnras},
         year = 2019,
        month = jun,
       volume = {486},
       number = {2},
        pages = {2827-2849},
          doi = {10.1093/mnras/stz937},
archivePrefix = {arXiv},
       eprint = {1901.10203},
 primaryClass = {astro-ph.GA},
       adsurl = {https://ui.adsabs.harvard.edu/abs/2019MNRAS.486.2827D}
}

@ARTICLE{vanderWel16,
       author = {{van der Wel}, A. and {Noeske}, K. and {Bezanson}, R. and {Pacifici}, C. and {Gallazzi}, A. and {Franx}, M. and {Mu{\~n}oz-Mateos}, J.~C. and {Bell}, E.~F. and {Brammer}, G. and {Charlot}, S. and {Chauk{\'e}}, P. and {Labb{\'e}}, I. and {Maseda}, M.~V. and {Muzzin}, A. and {Rix}, H.-W. and {Sobral}, D. and {van de Sande}, J. and {van Dokkum}, P.~G. and {Wild}, V. and {Wolf}, C.},
        title = "{The VLT LEGA-C Spectroscopic Survey: The Physics of Galaxies at a Lookback Time of 7 Gyr}",
      journal = {\apjs},
         year = 2016,
        month = apr,
       volume = {223},
       number = {2},
          eid = {29},
        pages = {29},
          doi = {10.3847/0067-0049/223/2/29},
archivePrefix = {arXiv},
       eprint = {1603.05479},
 primaryClass = {astro-ph.GA},
       adsurl = {https://ui.adsabs.harvard.edu/abs/2016ApJS..223...29V}
}

@ARTICLE{Loubser25,
       author = {{Loubser}, S. Ilani},
        title = "{Measuring the expansion history of the Universe with DESI cosmic chronometers}",
      journal = {\mnras},
         year = 2025,
        month = dec,
       volume = {544},
       number = {4},
        pages = {3064-3075},
          doi = {10.1093/mnras/staf1939},
archivePrefix = {arXiv},
       eprint = {2511.02730},
 primaryClass = {astro-ph.CO},
       adsurl = {https://ui.adsabs.harvard.edu/abs/2025MNRAS.544.3064L}
}

@ARTICLE{Tomasetti25,
       author = {{Tomasetti}, E. and {Moresco}, M. and {Granata}, G. and {D'Addona}, M. and {Bergamini}, P. and {Grillo}, C. and {Mercurio}, A. and {Rosati}, P. and {Cimatti}, A. and {Tortorelli}, L. and {Schuldt}, S. and {Meneghetti}, M.},
        title = "{Cosmic chronometers with galaxy clusters: A new avenue for multi-probe cosmology}",
      journal = {\aap},
         year = 2026,
        month = apr,
       volume = {708},
          eid = {A145},
        pages = {A145},
          doi = {10.1051/0004-6361/202558259},
archivePrefix = {arXiv},
       eprint = {2512.02109},
 primaryClass = {astro-ph.CO},
       adsurl = {https://ui.adsabs.harvard.edu/abs/2026A&A...708A.145T}
}

@ARTICLE{Zhang14,
       author = {{Zhang}, Cong and {Zhang}, Han and {Yuan}, Shuo and {Liu}, Siqi and {Zhang}, Tong-Jie and {Sun}, Yan-Chun},
        title = "{Four new observational H(z) data from luminous red galaxies in the Sloan Digital Sky Survey data release seven}",
      journal = {Research in Astronomy and Astrophysics},
         year = 2014,
        month = oct,
       volume = {14},
       number = {10},
          eid = {1221-1233},
        pages = {1221-1233},
          doi = {10.1088/1674-4527/14/10/002},
archivePrefix = {arXiv},
       eprint = {1207.4541},
 primaryClass = {astro-ph.CO},
       adsurl = {https://ui.adsabs.harvard.edu/abs/2014RAA....14.1221Z}
}

@ARTICLE{Moresco11,
       author = {{Moresco}, Michele and {Jimenez}, Raul and {Cimatti}, Andrea and {Pozzetti}, Lucia},
        title = "{Constraining the expansion rate of the Universe using low-redshift ellipticals as cosmic chronometers}",
      journal = {\jcap},
         year = 2011,
        month = mar,
       volume = {2011},
       number = {3},
          eid = {045},
        pages = {045},
          doi = {10.1088/1475-7516/2011/03/045},
archivePrefix = {arXiv},
       eprint = {1010.0831},
 primaryClass = {astro-ph.CO},
       adsurl = {https://ui.adsabs.harvard.edu/abs/2011JCAP...03..045M}
}

@ARTICLE{Thomas05,
       author = {{Thomas}, Daniel and {Maraston}, Claudia and {Bender}, Ralf and {Mendes de Oliveira}, Claudia},
        title = "{The Epochs of Early-Type Galaxy Formation as a Function of Environment}",
      journal = {\apj},
         year = 2005,
        month = mar,
       volume = {621},
       number = {2},
        pages = {673-694},
          doi = {10.1086/426932},
archivePrefix = {arXiv},
       eprint = {astro-ph/0410209},
 primaryClass = {astro-ph},
       adsurl = {https://ui.adsabs.harvard.edu/abs/2005ApJ...621..673T}
}

@ARTICLE{Trujillo14,
       author = {{Trujillo}, Ignacio and {Ferr{\'e}-Mateu}, Anna and {Balcells}, Marc and {Vazdekis}, Alexandre and {S{\'a}nchez-Bl{\'a}zquez}, Patricia},
        title = "{NGC 1277: A Massive Compact Relic Galaxy in the Nearby Universe}",
      journal = {\apjl},
         year = 2014,
        month = jan,
       volume = {780},
       number = {2},
          eid = {L20},
        pages = {L20},
          doi = {10.1088/2041-8205/780/2/L20},
archivePrefix = {arXiv},
       eprint = {1310.6367},
 primaryClass = {astro-ph.CO},
       adsurl = {https://ui.adsabs.harvard.edu/abs/2014ApJ...780L..20T}
}

@ARTICLE{Moresco16,
       author = {{Moresco}, Michele and {Pozzetti}, Lucia and {Cimatti}, Andrea and {Jimenez}, Raul and {Maraston}, Claudia and {Verde}, Licia and {Thomas}, Daniel and {Citro}, Annalisa and {Tojeiro}, Rita and {Wilkinson}, David},
        title = "{A 6\% measurement of the Hubble parameter at z\raisebox{-0.5ex}\textasciitilde0.45: direct evidence of the epoch of cosmic re-acceleration}",
      journal = {\jcap},
         year = 2016,
        month = may,
       volume = {2016},
       number = {5},
          eid = {014},
        pages = {014},
          doi = {10.1088/1475-7516/2016/05/014},
archivePrefix = {arXiv},
       eprint = {1601.01701},
 primaryClass = {astro-ph.CO},
       adsurl = {https://ui.adsabs.harvard.edu/abs/2016JCAP...05..014M}
}

@ARTICLE{Spiniello24,
       author = {{Spiniello}, C. and {D'Ago}, G. and {Coccato}, L. and {Hartke}, J. and {Tortora}, C. and {Ferr{\'e}-Mateu}, A. and {Pulsoni}, C. and {Cappellari}, M. and {Maksymowicz-Maciata}, M. and {Arnaboldi}, M. and {Bevacqua}, D. and {Gallazzi}, A. and {Hunt}, L.~K. and {La Barbera}, F. and {Mart{\'\i}n-Navarro}, I. and {Napolitano}, N.~R. and {Radovich}, M. and {Saracco}, P. and {Scognamiglio}, D. and {Spavone}, M. and {Zibetti}, S.},
        title = "{INSPIRE: INvestigating Stellar Population In RElics - V. A catalogue of ultra-compact massive galaxies outside the local Universe and their degree of relicness}",
      journal = {\mnras},
         year = 2024,
        month = jan,
       volume = {527},
       number = {3},
        pages = {8793-8811},
          doi = {10.1093/mnras/stad3703},
archivePrefix = {arXiv},
       eprint = {2309.12966},
 primaryClass = {astro-ph.GA},
       adsurl = {https://ui.adsabs.harvard.edu/abs/2024MNRAS.527.8793S}
}

@ARTICLE{labbe23,
       author = {{Labb{\'e}}, Ivo and {van Dokkum}, Pieter and {Nelson}, Erica and {Bezanson}, Rachel and {Suess}, Katherine A. and {Leja}, Joel and {Brammer}, Gabriel and {Whitaker}, Katherine and {Mathews}, Elijah and {Stefanon}, Mauro and {Wang}, Bingjie},
        title = "{A population of red candidate massive galaxies  600 Myr after the Big Bang}",
      journal = {\nat},
         year = 2023,
        month = apr,
       volume = {616},
       number = {7956},
        pages = {266-269},
          doi = {10.1038/s41586-023-05786-2},
archivePrefix = {arXiv},
       eprint = {2207.12446},
 primaryClass = {astro-ph.GA},
       adsurl = {https://ui.adsabs.harvard.edu/abs/2023Natur.616..266L}
}

@ARTICLE{Salim18,
       author = {{Salim}, Samir and {Boquien}, M{\'e}d{\'e}ric and {Lee}, Janice C.},
        title = "{Dust Attenuation Curves in the Local Universe: Demographics and New Laws for Star-forming Galaxies and High-redshift Analogs}",
      journal = {\apj},
         year = 2018,
        month = may,
       volume = {859},
       number = {1},
          eid = {11},
        pages = {11},
          doi = {10.3847/1538-4357/aabf3c},
archivePrefix = {arXiv},
       eprint = {1804.05850},
 primaryClass = {astro-ph.GA},
       adsurl = {https://ui.adsabs.harvard.edu/abs/2018ApJ...859...11S}
}

@ARTICLE{Baldry21,
       author = {{Baldry}, Ivan K. and {Sullivan}, Tricia and {Rani}, Raffaele and {Turner}, Sebastian},
        title = "{Compact galaxies and the size-mass galaxy distribution from a colour-selected sample at 0.04 < z < 0.15 supplemented by ugrizYJHK photometric redshifts}",
      journal = {\mnras},
         year = 2021,
        month = nov,
       volume = {500},
       number = {2},
        pages = {1557-1574},
          doi = {10.1093/mnras/staa3327},
archivePrefix = {arXiv},
       eprint = {2008.09625},
 primaryClass = {astro-ph.GA},
       adsurl = {https://ui.adsabs.harvard.edu/abs/2021MNRAS.500.1557B}
}

@ARTICLE{Spiniello21,
       author = {{Spiniello}, C. and {Tortora}, C. and {D'Ago}, G. and {Coccato}, L. and {La Barbera}, F. and {Ferr{\'e}-Mateu}, A. and {Napolitano}, N.~R. and {Spavone}, M. and {Scognamiglio}, D. and {Arnaboldi}, M. and {Gallazzi}, A. and {Hunt}, L. and {Moehler}, S. and {Radovich}, M. and {Zibetti}, S.},
        title = "{INSPIRE: INvestigating Stellar Population In RElics. I. Survey presentation and pilot study}",
      journal = {\aap},
         year = 2021,
        month = feb,
       volume = {646},
          eid = {A28},
        pages = {A28},
          doi = {10.1051/0004-6361/202038936},
archivePrefix = {arXiv},
       eprint = {2011.05347},
 primaryClass = {astro-ph.GA},
       adsurl = {https://ui.adsabs.harvard.edu/abs/2021A&A...646A..28S}
}

@ARTICLE{Balogh99,
       author = {{Balogh}, Michael L. and {Morris}, Simon L. and {Yee}, H.~K.~C. and {Carlberg}, R.~G. and {Ellingson}, Erica},
        title = "{Differential Galaxy Evolution in Cluster and Field Galaxies at z\raisebox{-0.5ex}\textasciitilde0.3}",
      journal = {\apj},
         year = 1999,
        month = dec,
       volume = {527},
       number = {1},
        pages = {54-79},
          doi = {10.1086/308056},
archivePrefix = {arXiv},
       eprint = {astro-ph/9906470},
 primaryClass = {astro-ph},
       adsurl = {https://ui.adsabs.harvard.edu/abs/1999ApJ...527...54B}
}

@ARTICLE{Bruzual83,
       author = {{Bruzual},A. G.},
        title = "{Spectral evolution of galaxies. I. Early-type systems.}",
      journal = {\apj},
         year = 1983,
        month = oct,
       volume = {273},
        pages = {105-127},
          doi = {10.1086/161352},
       adsurl = {https://ui.adsabs.harvard.edu/abs/1983ApJ...273..105B}
}

@ARTICLE{Vazdekis15,
       author = {{Vazdekis}, A. and {Coelho}, P. and {Cassisi}, S. and {Ricciardelli}, E. and {Falc{\'o}n-Barroso}, J. and {S{\'a}nchez-Bl{\'a}zquez}, P. and {La Barbera}, F. and {Beasley}, M.~A. and {Pietrinferni}, A.},
        title = "{Evolutionary stellar population synthesis with MILES - II. Scaled-solar and {\ensuremath{\alpha}}-enhanced models}",
      journal = {\mnras},
         year = 2015,
        month = may,
       volume = {449},
       number = {2},
        pages = {1177-1214},
          doi = {10.1093/mnras/stv151},
archivePrefix = {arXiv},
       eprint = {1504.08032},
 primaryClass = {astro-ph.GA},
       adsurl = {https://ui.adsabs.harvard.edu/abs/2015MNRAS.449.1177V}
}

@ARTICLE{Jimenez03,
       author = {{Jimenez}, Raul and {Verde}, Licia and {Treu}, Tommaso and {Stern}, Daniel},
        title = "{Constraints on the Equation of State of Dark Energy and the Hubble Constant from Stellar Ages and the Cosmic Microwave Background}",
      journal = {\apj},
         year = 2003,
        month = aug,
       volume = {593},
       number = {2},
        pages = {622-629},
          doi = {10.1086/376595},
archivePrefix = {arXiv},
       eprint = {astro-ph/0302560},
 primaryClass = {astro-ph},
       adsurl = {https://ui.adsabs.harvard.edu/abs/2003ApJ...593..622J}
}

@ARTICLE{Simon05,
       author = {{Simon}, Joan and {Verde}, Licia and {Jimenez}, Raul},
        title = "{Constraints on the redshift dependence of the dark energy potential}",
      journal = {\prd},
         year = 2005,
        month = jun,
       volume = {71},
       number = {12},
          eid = {123001},
        pages = {123001},
          doi = {10.1103/PhysRevD.71.123001},
archivePrefix = {arXiv},
       eprint = {astro-ph/0412269},
 primaryClass = {astro-ph},
       adsurl = {https://ui.adsabs.harvard.edu/abs/2005PhRvD..71l3001S}
}

@ARTICLE{Stern10,
       author = {{Stern}, Daniel and {Jimenez}, Raul and {Verde}, Licia and {Kamionkowski}, Marc and {Stanford}, S. Adam},
        title = "{Cosmic chronometers: constraining the equation of state of dark energy. I: H(z) measurements}",
      journal = {\jcap},
         year = 2010,
        month = feb,
       volume = {2010},
       number = {2},
          eid = {008},
        pages = {008},
          doi = {10.1088/1475-7516/2010/02/008},
archivePrefix = {arXiv},
       eprint = {0907.3149},
 primaryClass = {astro-ph.CO},
       adsurl = {https://ui.adsabs.harvard.edu/abs/2010JCAP...02..008S}
}

@INPROCEEDINGS{Stoehr08,
       author = {{Stoehr}, F. and {White}, R. and {Smith}, M. and {Kamp}, I. and {Thompson}, R. and {Durand}, D. and {Freudling}, W. and {Fraquelli}, D. and {Haase}, J. and {Hook}, R. and {Kimball}, T. and {K{\"u}mmel}, M. and {Levay}, K. and {Lombardi}, M. and {Micol}, A. and {Rogers}, T.},
        title = "{DER\_SNR: A Simple \& General Spectroscopic Signal-to-Noise Measurement Algorithm}",
    booktitle = {Astronomical Data Analysis Software and Systems XVII},
         year = 2008,
       editor = {{Argyle}, R.~W. and {Bunclark}, P.~S. and {Lewis}, J.~R.},
       series = {Astronomical Society of the Pacific Conference Series},
       volume = {394},
        month = aug,
        pages = {505},
       adsurl = {https://ui.adsabs.harvard.edu/abs/2008ASPC..394..505S}
}

@ARTICLE{Lisiecki23,
       author = {{Lisiecki}, Krzysztof and {Ma{\l}ek}, Katarzyna and {Siudek}, Ma{\l}gorzata and {Pollo}, Agnieszka and {Krywult}, Janusz and {Karska}, Agata and {Junais}},
        title = "{The first catalogue of spectroscopically confirmed red nuggets at z {\ensuremath{\sim}} 0.7 from the VIPERS survey. Linking high-z red nuggets and local relics}",
      journal = {\aap},
         year = 2023,
        month = jan,
       volume = {669},
          eid = {A95},
        pages = {A95},
          doi = {10.1051/0004-6361/202243616},
archivePrefix = {arXiv},
       eprint = {2208.04601},
 primaryClass = {astro-ph.GA},
       adsurl = {https://ui.adsabs.harvard.edu/abs/2023A&A...669A..95L}
}

@ARTICLE{Almeida23,
       author = {{Almeida}, Andr{\'e}s and {Anderson}, Scott F. and {Argudo-Fern{\'a}ndez}, Maria and {Badenes}, Carles and {Barger}, Kat and {Barrera-Ballesteros}, Jorge K. and {Bender}, Chad F. and {Benitez}, Erika and {Besser}, Felipe and {Bird}, Jonathan C. and {Bizyaev}, Dmitry and {Blanton}, Michael R. and {Bochanski}, John and {Bovy}, Jo and {Brandt}, William Nielsen and {Brownstein}, Joel R. and {Buchner}, Johannes and {Bulbul}, Esra and {Burchett}, Joseph N. and {Cano D{\'\i}az}, Mariana and {Carlberg}, Joleen K. and {Casey}, Andrew R. and {Chandra}, Vedant and {Cherinka}, Brian and {Chiappini}, Cristina and {Coker}, Abigail A. and {Comparat}, Johan and {Conroy}, Charlie and {Contardo}, Gabriella and {Cortes}, Arlin and {Covey}, Kevin and {Crane}, Jeffrey D. and {Cunha}, Katia and {Dabbieri}, Collin and {Davidson}, James W. and {Davis}, Megan C. and {de Andrade Queiroz}, Anna Barbara and {De Lee}, Nathan and {M{\'e}ndez Delgado}, Jos{\'e} Eduardo and {Demasi}, Sebastian and {Di Mille}, Francesco and {Donor}, John and {Dow}, Peter and {Dwelly}, Tom and {Eracleous}, Mike and {Eriksen}, Jamey and {Fan}, Xiaohui and {Farr}, Emily and {Frederick}, Sara and {Fries}, Logan and {Frinchaboy}, Peter and {G{\"a}nsicke}, Boris T. and {Ge}, Junqiang and {Gonz{\'a}lez {\'A}vila}, Consuelo and {Grabowski}, Katie and {Grier}, Catherine and {Guiglion}, Guillaume and {Gupta}, Pramod and {Hall}, Patrick and {Hawkins}, Keith and {Hayes}, Christian R. and {Hermes}, J.~J. and {Hern{\'a}ndez-Garc{\'\i}a}, Lorena and {Hogg}, David W. and {Holtzman}, Jon A. and {Ibarra-Medel}, Hector Javier and {Ji}, Alexander and {Jofre}, Paula and {Johnson}, Jennifer A. and {Jones}, Amy M. and {Kinemuchi}, Karen and {Kluge}, Matthias and {Koekemoer}, Anton and {Kollmeier}, Juna A. and {Kounkel}, Marina and {Krishnarao}, Dhanesh and {Krumpe}, Mirko and {Lacerna}, Ivan and {Lago}, Paulo Jakson Assuncao and {Laporte}, Chervin and {Liu}, Chao and {Liu}, Ang and {Liu}, Xin and {Lopes}, Alexandre Roman and {Macktoobian}, Matin and {Majewski}, Steven R. and {Malanushenko}, Viktor and {Maoz}, Dan and {Masseron}, Thomas and {Masters}, Karen L. and {Matijevic}, Gal and {McBride}, Aidan and {Medan}, Ilija and {Merloni}, Andrea and {Morrison}, Sean and {Myers}, Natalie and {M{\'e}sz{\'a}ros}, Szabolcs and {Negrete}, C. Alenka and {Nidever}, David L. and {Nitschelm}, Christian and {Oravetz}, Daniel and {Oravetz}, Audrey and {Pan}, Kaike and {Peng}, Yingjie and {Pinsonneault}, Marc H. and {Pogge}, Rick and {Qiu}, Dan and {Ramirez}, Solange V. and {Rix}, Hans-Walter and {Fern{\'a}ndez Rosso}, Daniela and {Runnoe}, Jessie and {Salvato}, Mara and {Sanchez}, Sebastian F. and {Santana}, Felipe A. and {Saydjari}, Andrew and {Sayres}, Conor and {Schlaufman}, Kevin C. and {Schneider}, Donald P. and {Schwope}, Axel and {Serna}, Javier and {Shen}, Yue and {Sobeck}, Jennifer and {Song}, Ying-Yi and {Souto}, Diogo and {Spoo}, Taylor and {Stassun}, Keivan G. and {Steinmetz}, Matthias and {Straumit}, Ilya and {Stringfellow}, Guy and {S{\'a}nchez-Gallego}, Jos{\'e} and {Taghizadeh-Popp}, Manuchehr and {Tayar}, Jamie and {Thakar}, Ani and {Tissera}, Patricia B. and {Tkachenko}, Andrew and {Hernandez Toledo}, Hector and {Trakhtenbrot}, Benny and {Fern{\'a}ndez-Trincado}, Jos{\'e} G. and {Troup}, Nicholas and {Trump}, Jonathan R. and {Tuttle}, Sarah and {Ulloa}, Natalie and {Vazquez-Mata}, Jose Antonio and {Vera Alfaro}, Pablo and {Villanova}, Sandro and {Wachter}, Stefanie and {Weijmans}, Anne-Marie and {Wheeler}, Adam and {Wilson}, John and {Wojno}, Leigh and {Wolf}, Julien and {Xue}, Xiang-Xiang and {Ybarra}, Jason E. and {Zari}, Eleonora and {Zasowski}, Gail},
        title = "{The Eighteenth Data Release of the Sloan Digital Sky Surveys: Targeting and First Spectra from SDSS-V}",
      journal = {\apjs},
         year = 2023,
        month = aug,
       volume = {267},
       number = {2},
          eid = {44},
        pages = {44},
          doi = {10.3847/1538-4365/acda98},
archivePrefix = {arXiv},
       eprint = {2301.07688},
 primaryClass = {astro-ph.GA},
       adsurl = {https://ui.adsabs.harvard.edu/abs/2023ApJS..267...44A}
}

@ARTICLE{vandokkum10,
       author = {{van Dokkum}, Pieter G. and {Whitaker}, Katherine E. and {Brammer}, Gabriel and {Franx}, Marijn and {Kriek}, Mariska and {Labb{\'e}}, Ivo and {Marchesini}, Danilo and {Quadri}, Ryan and {Bezanson}, Rachel and {Illingworth}, Garth D. and {Muzzin}, Adam and {Rudnick}, Gregory and {Tal}, Tomer and {Wake}, David},
        title = "{The Growth of Massive Galaxies Since z = 2}",
      journal = {\apj},
         year = 2010,
        month = feb,
       volume = {709},
       number = {2},
        pages = {1018-1041},
          doi = {10.1088/0004-637X/709/2/1018},
archivePrefix = {arXiv},
       eprint = {0912.0514},
 primaryClass = {astro-ph.CO},
       adsurl = {https://ui.adsabs.harvard.edu/abs/2010ApJ...709.1018V}
}

@ARTICLE{oser10,
       author = {{Oser}, Ludwig and {Ostriker}, Jeremiah P. and {Naab}, Thorsten and {Johansson}, Peter H. and {Burkert}, Andreas},
        title = "{The Two Phases of Galaxy Formation}",
      journal = {\apj},
         year = 2010,
        month = dec,
       volume = {725},
       number = {2},
        pages = {2312-2323},
          doi = {10.1088/0004-637X/725/2/2312},
archivePrefix = {arXiv},
       eprint = {1010.1381},
 primaryClass = {astro-ph.CO},
       adsurl = {https://ui.adsabs.harvard.edu/abs/2010ApJ...725.2312O}
}

@ARTICLE{flores-freitas21,
       author = {{Flores-Freitas}, Rodrigo and {Chies-Santos}, Ana L. and {Furlanetto}, Cristina and {De Rossi}, Mar{\'\i}a Emilia and {Ferreira}, Leonardo and {Zenocratti}, Lucas J. and {Alamo-Mart{\'\i}nez}, Karla A.},
        title = "{Relic galaxy analogues in TNG50 simulation: the formation pathways of surviving red nuggets in a cosmological simulation}",
      journal = {\mnras},
         year = 2022,
        month = may,
       volume = {512},
       number = {1},
        pages = {245-264},
          doi = {10.1093/mnras/stac187},
archivePrefix = {arXiv},
       eprint = {2112.12846},
 primaryClass = {astro-ph.GA},
       adsurl = {https://ui.adsabs.harvard.edu/abs/2022MNRAS.512..245F}
}

@ARTICLE{zolotov15,
       author = {{Zolotov}, Adi and {Dekel}, Avishai and {Mandelker}, Nir and {Tweed}, Dylan and {Inoue}, Shigeki and {DeGraf}, Colin and {Ceverino}, Daniel and {Primack}, Joel R. and {Barro}, Guillermo and {Faber}, Sandra M.},
        title = "{Compaction and quenching of high-z galaxies in cosmological simulations: blue and red nuggets}",
      journal = {\mnras},
         year = 2015,
        month = jul,
       volume = {450},
       number = {3},
        pages = {2327-2353},
          doi = {10.1093/mnras/stv740},
archivePrefix = {arXiv},
       eprint = {1412.4783},
 primaryClass = {astro-ph.GA},
       adsurl = {https://ui.adsabs.harvard.edu/abs/2015MNRAS.450.2327Z}
}

@ARTICLE{naab09,
       author = {{Naab}, Thorsten and {Johansson}, Peter H. and {Ostriker}, Jeremiah P.},
        title = "{Minor Mergers and the Size Evolution of Elliptical Galaxies}",
      journal = {\apjl},
         year = 2009,
        month = jul,
       volume = {699},
       number = {2},
        pages = {L178-L182},
          doi = {10.1088/0004-637X/699/2/L178},
archivePrefix = {arXiv},
       eprint = {0903.1636},
 primaryClass = {astro-ph.CO},
       adsurl = {https://ui.adsabs.harvard.edu/abs/2009ApJ...699L.178N}
}

@ARTICLE{Planck,
       author = {{Planck Collaboration} and {Aghanim}, N. and {Akrami}, Y. and {Ashdown}, M. and {Aumont}, J. and {Baccigalupi}, C. and {Ballardini}, M. and {Banday}, A.~J. and {Barreiro}, R.~B. and {Bartolo}, N. and {Basak}, S. and {Battye}, R. and {Benabed}, K. and {Bernard}, J.-P. and {Bersanelli}, M. and {Bielewicz}, P. and {Bock}, J.~J. and {Bond}, J.~R. and {Borrill}, J. and {Bouchet}, F.~R. and {Boulanger}, F. and {Bucher}, M. and {Burigana}, C. and {Butler}, R.~C. and {Calabrese}, E. and {Cardoso}, J.-F. and {Carron}, J. and {Challinor}, A. and {Chiang}, H.~C. and {Chluba}, J. and {Colombo}, L.~P.~L. and {Combet}, C. and {Contreras}, D. and {Crill}, B.~P. and {Cuttaia}, F. and {de Bernardis}, P. and {de Zotti}, G. and {Delabrouille}, J. and {Delouis}, J.-M. and {Di Valentino}, E. and {Diego}, J.~M. and {Dor{\'e}}, O. and {Douspis}, M. and {Ducout}, A. and {Dupac}, X. and {Dusini}, S. and {Efstathiou}, G. and {Elsner}, F. and {En{\ss}lin}, T.~A. and {Eriksen}, H.~K. and {Fantaye}, Y. and {Farhang}, M. and {Fergusson}, J. and {Fernandez-Cobos}, R. and {Finelli}, F. and {Forastieri}, F. and {Frailis}, M. and {Fraisse}, A.~A. and {Franceschi}, E. and {Frolov}, A. and {Galeotta}, S. and {Galli}, S. and {Ganga}, K. and {G{\'e}nova-Santos}, R.~T. and {Gerbino}, M. and {Ghosh}, T. and {Gonz{\'a}lez-Nuevo}, J. and {G{\'o}rski}, K.~M. and {Gratton}, S. and {Gruppuso}, A. and {Gudmundsson}, J.~E. and {Hamann}, J. and {Handley}, W. and {Hansen}, F.~K. and {Herranz}, D. and {Hildebrandt}, S.~R. and {Hivon}, E. and {Huang}, Z. and {Jaffe}, A.~H. and {Jones}, W.~C. and {Karakci}, A. and {Keih{\"a}nen}, E. and {Keskitalo}, R. and {Kiiveri}, K. and {Kim}, J. and {Kisner}, T.~S. and {Knox}, L. and {Krachmalnicoff}, N. and {Kunz}, M. and {Kurki-Suonio}, H. and {Lagache}, G. and {Lamarre}, J.-M. and {Lasenby}, A. and {Lattanzi}, M. and {Lawrence}, C.~R. and {Le Jeune}, M. and {Lemos}, P. and {Lesgourgues}, J. and {Levrier}, F. and {Lewis}, A. and {Liguori}, M. and {Lilje}, P.~B. and {Lilley}, M. and {Lindholm}, V. and {L{\'o}pez-Caniego}, M. and {Lubin}, P.~M. and {Ma}, Y.-Z. and {Mac{\'\i}as-P{\'e}rez}, J.~F. and {Maggio}, G. and {Maino}, D. and {Mandolesi}, N. and {Mangilli}, A. and {Marcos-Caballero}, A. and {Maris}, M. and {Martin}, P.~G. and {Martinelli}, M. and {Mart{\'\i}nez-Gonz{\'a}lez}, E. and {Matarrese}, S. and {Mauri}, N. and {McEwen}, J.~D. and {Meinhold}, P.~R. and {Melchiorri}, A. and {Mennella}, A. and {Migliaccio}, M. and {Millea}, M. and {Mitra}, S. and {Miville-Desch{\^e}nes}, M.-A. and {Molinari}, D. and {Montier}, L. and {Morgante}, G. and {Moss}, A. and {Natoli}, P. and {N{\o}rgaard-Nielsen}, H.~U. and {Pagano}, L. and {Paoletti}, D. and {Partridge}, B. and {Patanchon}, G. and {Peiris}, H.~V. and {Perrotta}, F. and {Pettorino}, V. and {Piacentini}, F. and {Polastri}, L. and {Polenta}, G. and {Puget}, J.-L. and {Rachen}, J.~P. and {Reinecke}, M. and {Remazeilles}, M. and {Renzi}, A. and {Rocha}, G. and {Rosset}, C. and {Roudier}, G. and {Rubi{\~n}o-Mart{\'\i}n}, J.~A. and {Ruiz-Granados}, B. and {Salvati}, L. and {Sandri}, M. and {Savelainen}, M. and {Scott}, D. and {Shellard}, E.~P.~S. and {Sirignano}, C. and {Sirri}, G. and {Spencer}, L.~D. and {Sunyaev}, R. and {Suur-Uski}, A.-S. and {Tauber}, J.~A. and {Tavagnacco}, D. and {Tenti}, M. and {Toffolatti}, L. and {Tomasi}, M. and {Trombetti}, T. and {Valenziano}, L. and {Valiviita}, J. and {Van Tent}, B. and {Vibert}, L. and {Vielva}, P. and {Villa}, F. and {Vittorio}, N. and {Wandelt}, B.~D. and {Wehus}, I.~K. and {White}, M. and {White}, S.~D.~M. and {Zacchei}, A. and {Zonca}, A.},
        title = "{Planck 2018 results. VI. Cosmological parameters}",
      journal = {\aap},
         year = 2020,
        month = sep,
       volume = {641},
          eid = {A6},
        pages = {A6},
          doi = {10.1051/0004-6361/201833910},
archivePrefix = {arXiv},
       eprint = {1807.06209},
 primaryClass = {astro-ph.CO},
       adsurl = {https://ui.adsabs.harvard.edu/abs/2020A&A...641A...6P}
}

@ARTICLE{Bertone18,
       author = {{Bertone}, Gianfranco and {Hooper}, Dan},
        title = "{History of dark matter}",
      journal = {Reviews of Modern Physics},
         year = 2018,
        month = oct,
       volume = {90},
       number = {4},
          eid = {045002},
        pages = {045002},
          doi = {10.1103/RevModPhys.90.045002},
archivePrefix = {arXiv},
       eprint = {1605.04909},
 primaryClass = {astro-ph.CO},
       adsurl = {https://ui.adsabs.harvard.edu/abs/2018RvMP...90d5002B}
}

@ARTICLE{Worthey94,
       author = {{Worthey}, Guy},
        title = "{Comprehensive Stellar Population Models and the Disentanglement of Age and Metallicity Effects}",
      journal = {\apjs},
         year = 1994,
        month = nov,
       volume = {95},
        pages = {107},
          doi = {10.1086/192096},
       adsurl = {https://ui.adsabs.harvard.edu/abs/1994ApJS...95..107W}
}

@ARTICLE{Kaviraj07,
       author = {{Kaviraj}, S. and {Rey}, S.-C. and {Rich}, R.~M. and {Yoon}, S.-J. and {Yi}, S.~K.},
        title = "{Better age estimation using ultraviolet-optical colours: breaking the age-metallicity degeneracy}",
      journal = {\mnras},
         year = 2007,
        month = oct,
       volume = {381},
       number = {1},
        pages = {L74-L78},
          doi = {10.1111/j.1745-3933.2007.00370.x},
archivePrefix = {arXiv},
       eprint = {astro-ph/0601050},
 primaryClass = {astro-ph},
       adsurl = {https://ui.adsabs.harvard.edu/abs/2007MNRAS.381L..74K}
}

@ARTICLE{Thomas99,
       author = {{Thomas}, D. and {Greggio}, L. and {Bender}, R.},
        title = "{Constraints on galaxy formation from alpha-enhancement in luminous elliptical galaxies}",
      journal = {\mnras},
         year = 1999,
        month = jan,
       volume = {302},
       number = {3},
        pages = {537-548},
          doi = {10.1046/j.1365-8711.1999.02138.x},
archivePrefix = {arXiv},
       eprint = {astro-ph/9809261},
 primaryClass = {astro-ph},
       adsurl = {https://ui.adsabs.harvard.edu/abs/1999MNRAS.302..537T}
}

@ARTICLE{Maoz10,
       author = {{Maoz}, Dan and {Sharon}, Keren and {Gal-Yam}, Avishay},
        title = "{The Supernova Delay Time Distribution in Galaxy Clusters and Implications for Type-Ia Progenitors and Metal Enrichment}",
      journal = {\apj},
         year = 2010,
        month = oct,
       volume = {722},
       number = {2},
        pages = {1879-1894},
          doi = {10.1088/0004-637X/722/2/1879},
archivePrefix = {arXiv},
       eprint = {1006.3576},
 primaryClass = {astro-ph.CO},
       adsurl = {https://ui.adsabs.harvard.edu/abs/2010ApJ...722.1879M}
}

@ARTICLE{Walcher15,
       author = {{Walcher}, C.~J. and {Coelho}, P.~R.~T. and {Gallazzi}, A. and {Bruzual}, G. and {Charlot}, S. and {Chiappini}, C.},
        title = "{Abundance patterns in early-type galaxies: is there a ``knee'' in the [Fe/H] vs. [{\ensuremath{\alpha}}/Fe] relation?}",
      journal = {\aap},
         year = 2015,
        month = oct,
       volume = {582},
          eid = {A46},
        pages = {A46},
          doi = {10.1051/0004-6361/201525924},
archivePrefix = {arXiv},
       eprint = {1508.05103},
 primaryClass = {astro-ph.GA},
       adsurl = {https://ui.adsabs.harvard.edu/abs/2015A&A...582A..46W}
}

@ARTICLE{Alam17,
       author = {{Alam}, Shadab and {Ata}, Metin and {Bailey}, Stephen and {Beutler}, Florian and {Bizyaev}, Dmitry and {Blazek}, Jonathan A. and {Bolton}, Adam S. and {Brownstein}, Joel R. and {Burden}, Angela and {Chuang}, Chia-Hsun and {Comparat}, Johan and {Cuesta}, Antonio J. and {Dawson}, Kyle S. and {Eisenstein}, Daniel J. and {Escoffier}, Stephanie and {Gil-Mar{\'\i}n}, H{\'e}ctor and {Grieb}, Jan Niklas and {Hand}, Nick and {Ho}, Shirley and {Kinemuchi}, Karen and {Kirkby}, David and {Kitaura}, Francisco and {Malanushenko}, Elena and {Malanushenko}, Viktor and {Maraston}, Claudia and {McBride}, Cameron K. and {Nichol}, Robert C. and {Olmstead}, Matthew D. and {Oravetz}, Daniel and {Padmanabhan}, Nikhil and {Palanque-Delabrouille}, Nathalie and {Pan}, Kaike and {Pellejero-Ibanez}, Marcos and {Percival}, Will J. and {Petitjean}, Patrick and {Prada}, Francisco and {Price-Whelan}, Adrian M. and {Reid}, Beth A. and {Rodr{\'\i}guez-Torres}, Sergio A. and {Roe}, Natalie A. and {Ross}, Ashley J. and {Ross}, Nicholas P. and {Rossi}, Graziano and {Rubi{\~n}o-Mart{\'\i}n}, Jose Alberto and {Saito}, Shun and {Salazar-Albornoz}, Salvador and {Samushia}, Lado and {S{\'a}nchez}, Ariel G. and {Satpathy}, Siddharth and {Schlegel}, David J. and {Schneider}, Donald P. and {Sc{\'o}ccola}, Claudia G. and {Seo}, Hee-Jong and {Sheldon}, Erin S. and {Simmons}, Audrey and {Slosar}, An{\v{z}}e and {Strauss}, Michael A. and {Swanson}, Molly E.~C. and {Thomas}, Daniel and {Tinker}, Jeremy L. and {Tojeiro}, Rita and {Maga{\~n}a}, Mariana Vargas and {Vazquez}, Jose Alberto and {Verde}, Licia and {Wake}, David A. and {Wang}, Yuting and {Weinberg}, David H. and {White}, Martin and {Wood-Vasey}, W. Michael and {Y{\`e}che}, Christophe and {Zehavi}, Idit and {Zhai}, Zhongxu and {Zhao}, Gong-Bo},
        title = "{The clustering of galaxies in the completed SDSS-III Baryon Oscillation Spectroscopic Survey: cosmological analysis of the DR12 galaxy sample}",
      journal = {\mnras},
         year = 2017,
        month = sep,
       volume = {470},
       number = {3},
        pages = {2617-2652},
          doi = {10.1093/mnras/stx721},
archivePrefix = {arXiv},
       eprint = {1607.03155},
 primaryClass = {astro-ph.CO},
       adsurl = {https://ui.adsabs.harvard.edu/abs/2017MNRAS.470.2617A}
}

@ARTICLE{DiValentino,
       author = {{Di Valentino}, Eleonora and {Mena}, Olga and {Pan}, Supriya and {Visinelli}, Luca and {Yang}, Weiqiang and {Melchiorri}, Alessandro and {Mota}, David F. and {Riess}, Adam G. and {Silk}, Joseph},
        title = "{In the realm of the Hubble tension-a review of solutions}",
      journal = {Classical and Quantum Gravity},
         year = 2021,
        month = jul,
       volume = {38},
       number = {15},
          eid = {153001},
        pages = {153001},
          doi = {10.1088/1361-6382/ac086d},
archivePrefix = {arXiv},
       eprint = {2103.01183},
 primaryClass = {astro-ph.CO},
       adsurl = {https://ui.adsabs.harvard.edu/abs/2021CQGra..38o3001D}
}

@ARTICLE{Hu_H0,
       author = {{Hu}, Jian-Ping and {Wang}, Fa-Yin},
        title = "{Hubble Tension: The Evidence of New Physics}",
      journal = {Universe},
         year = 2023,
        month = feb,
       volume = {9},
       number = {2},
          eid = {94},
        pages = {94},
          doi = {10.3390/universe9020094},
archivePrefix = {arXiv},
       eprint = {2302.05709},
 primaryClass = {astro-ph.CO},
       adsurl = {https://ui.adsabs.harvard.edu/abs/2023Univ....9...94H}
}

@ARTICLE{Verde,
       author = {{Verde}, Licia and {Treu}, Tommaso and {Riess}, Adam G.},
        title = "{Tensions between the early and late Universe}",
      journal = {Nature Astronomy},
         year = 2019,
        month = sep,
       volume = {3},
        pages = {891-895},
          doi = {10.1038/s41550-019-0902-0},
archivePrefix = {arXiv},
       eprint = {1907.10625},
 primaryClass = {astro-ph.CO},
       adsurl = {https://ui.adsabs.harvard.edu/abs/2019NatAs...3..891V}
}

@ARTICLE{Morescoreview,
       author = {{Moresco}, Michele and {Amati}, Lorenzo and {Amendola}, Luca and {Birrer}, Simon and {Blakeslee}, John P. and {Cantiello}, Michele and {Cimatti}, Andrea and {Darling}, Jeremy and {Della Valle}, Massimo and {Fishbach}, Maya and {Grillo}, Claudio and {Hamaus}, Nico and {Holz}, Daniel and {Izzo}, Luca and {Jimenez}, Raul and {Lusso}, Elisabeta and {Meneghetti}, Massimo and {Piedipalumbo}, Ester and {Pisani}, Alice and {Pourtsidou}, Alkistis and {Pozzetti}, Lucia and {Quartin}, Miguel and {Risaliti}, Guido and {Rosati}, Piero and {Verde}, Licia},
        title = "{Unveiling the Universe with emerging cosmological probes}",
      journal = {Living Reviews in Relativity},
         year = 2022,
        month = dec,
       volume = {25},
       number = {1},
          eid = {6},
        pages = {6},
          doi = {10.1007/s41114-022-00040-z},
archivePrefix = {arXiv},
       eprint = {2201.07241},
 primaryClass = {astro-ph.CO},
       adsurl = {https://ui.adsabs.harvard.edu/abs/2022LRR....25....6M}
}

@ARTICLE{JimenezFirst,
       author = {{Jimenez}, Raul and {Loeb}, Abraham},
        title = "{Constraining Cosmological Parameters Based on Relative Galaxy Ages}",
      journal = {\apj},
         year = 2002,
        month = jul,
       volume = {573},
       number = {1},
        pages = {37-42},
          doi = {10.1086/340549},
archivePrefix = {arXiv},
       eprint = {astro-ph/0106145},
 primaryClass = {astro-ph},
       adsurl = {https://ui.adsabs.harvard.edu/abs/2002ApJ...573...37J}
}

\begin{appendix}
\section{Cosmological model--DoR dependence}\label{app:cosmodep}

To quantify the dependence of the relic selection on the assumed cosmological model, we repeated the full {\tt pPXF} analysis using the pipeline described by \citet{Mills25} and \citet{Rosen+26}. 
We considered three commonly adopted flat $\Lambda$CDM cosmologies:
\begin{itemize}
    \item $H_0=67.7\textrm{ km\,s}^{-1}\,\textrm{Mpc}^{-1}$, $\Omega_\Lambda=0.689$ and $\Omega_\textrm{M}=0.311$ \citep[][]{planck16};
    \item $H_0=70.4\textrm{ km\,s}^{-1}\,\textrm{Mpc}^{-1}$, $\Omega_\Lambda=0.728$ and $\Omega_\textrm{M}=0.272$ \citep[WMAP7;][]{Komatsu11}
    \item $H_0=73.2\textrm{ km\,s}^{-1}\,\textrm{Mpc}^{-1}$, $\Omega_\Lambda=0.761$ and $\Omega_\textrm{M}=0.241$ \citep[WMAP3;][]{Spergel07}.
\end{itemize}
For computational efficiency, we repeated the analysis only for galaxies satisfying the initial selection criteria ($0.07 \leq z \leq 0.22$, $M_\star \geq 10^{10.8}\,M_\odot$, and $\sigma_{\rm DoR}<0.2$). For each cosmological model, we recomputed the SFHs and the corresponding DoR values before applying the final relic selection.
The resulting DoR distributions are shown in Figure~\ref{fig:comp_cosmo}. The distributions are nearly indistinguishable, although individual galaxies exhibit small shifts in their DoR values. 
In the most extreme comparison (Planck versus WMAP3), two galaxies leave the final sample while two different galaxies enter it. 
Since the final selection is based on a percentile threshold in the DoR distribution, the total sample size remains unchanged.
Finally, we repeated the $D_n4000$--redshift analysis for the three cosmological models and found that the recovered slope and its uncertainty vary by an amount that is negligible compared to the statistical uncertainties:
\begin{equation}
    \frac{\mbox{d}D_n4000}{\mbox{d}z} [\textrm{WMAP3]}= -0.34 \pm 0.18 \textrm{ [fit] } \pm 0.08 \textrm{ [bin] }.
\end{equation}
We therefore conclude that, although the adopted cosmology formally affects the reconstructed SFHs and the derived DoR values, its impact on the final selection and the inferred $H(z)$ measurement is negligible for the purposes of this work.

\begin{figure}[ht]
\centering
\includegraphics[width = 0.48\textwidth]{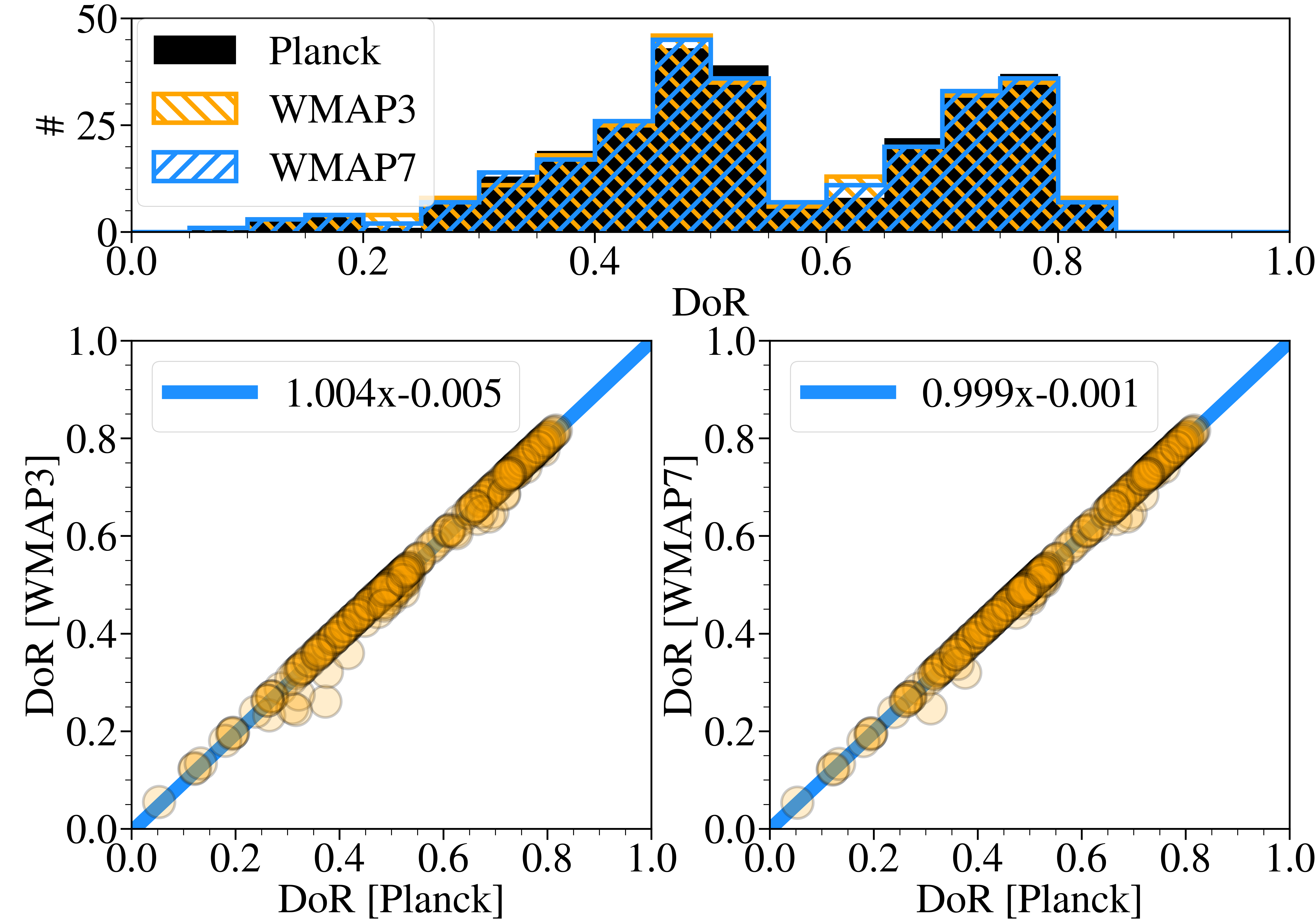}
\caption{
Comparison of the DoR obtained assuming three different flat $\Lambda$CDM cosmologies.
\textit{Top:} Distribution of DoR values for the three cosmological models.
\textit{Left:} Galaxy-by-galaxy comparison of the DoR values obtained assuming the Planck cosmology with those obtained using WMAP3, (\textit{right:}) and WMAP7 (\textit{right}). The blue solid lines indicate the best-fitting linear relation.}
\label{fig:comp_cosmo}
\end{figure}

\section{Traditional CC selection}\label{app:traditonal}

The selection of massive, passively evolving galaxies as CC is typically based on a combination of photometric and spectroscopic criteria. 
The exact implementation varies between studies depending on the desired balance between sample completeness and purity. 
Here, we investigate what fraction of our final sample would also satisfy the traditional CC selection criteria.

We begin with the photometric $UVJ$ diagnostic.
It is used to distinguish dusty star-forming galaxies from genuinely quiescent galaxies by breaking the degeneracy between dust reddening and old stellar populations \citet[][see also \citealt{Moresco18} and \citealt{Borghi22}, for NUV$rJ$ selection]{Tomasetti23}.
We use the \cite{Williams09} prescription for $UVJ$ in which $(U-V)>1.3$ and $(V-J)>1.6$ and $(U-V)>0.88\times(V-J)+0.69$ as E-INSPIRE reaches to $z<0.3$. 
We compile photometric observations from the SDSS ($ugriz$) \citep{Almeida23}, 2MASS ($JHK$) \citep{Skrutskie06}, and WISE ($W1$ and $W2$) \citep{Cutri12} surveys. 
Observed magnitudes are first corrected for Galactic extinction using the \texttt{dustmaps} package \citep{Green18}. 
Then extinction-corrected photometry is K-corrected and converted into the rest-frame using \texttt{EAZY} \citep{Brammer08}. 
The resulting UVJ diagram is shown in Figure.~\ref{fig:traditionalCC}.

We then consider four spectroscopic diagnostics commonly adopted in CC studies:
(i) [OII] $\lambda3727$ emission,
(ii) H$\alpha$ emission,
(iii) H$\delta$ absorption,
and (iv) the Ca II H:K ratio.
Detectable [OII] and H$\alpha$ emission are indicative of stellar populations in which star formation is recent or ongoing \citep[e.g.][]{Kewley19} whereas H$\delta$ absorption indicates a recent but since halted starburst event \citep{Charlot94, Pimbblet19}.
The H:K ratio provides a method of estimating whether a galaxy is contaminated by a younger stellar population \citep{Rose84, Pimbblet19}.

We use SDSS spectra prepared by the original E-INSPIRE survey.
Equivalent width (EW) measurements are found using an index-based pseudo-continuum approach similar to that used for calculating Lick indices. 
We calculate the mean flux density within the blue and red continuum windows for a given feature, defined in Table~\ref{tab:bandpasses}, and linearly interpolate across each feature bandpass to estimate the local continuum. 
We calculate the EW in the feature bandpass defined by $\lambda_1$ and $\lambda_2$ for each feature as:
\begin{equation}
EW = \int_{\lambda_1}^{\lambda_2}\left(1-\frac{F_\lambda}{F_{\mathrm{c}}(\lambda)}\right) d\lambda
\end{equation}
where $F_\lambda$ is the observed flux density, $F_{\mathrm{c}}(\lambda)$ is the interpolated pseudo-continuum. 
Prior to measurement, each spectrum is linearly interpolated onto a uniform wavelength grid spanning the full extent of the index definition to ensure consistent sampling of the bandpasses. 
Pixels flagged as unreliable by the spectral quality mask were excluded from the continuum and feature measurements, and the fraction of masked pixels within the combined bandpasses is recorded as a bad-pixel ratio to assess the reliability of each index measurement. 
Not all spectra provide reliable measurements in every wavelength region. 
Whenever the spectral quality is insufficient to measure a given diagnostic, the corresponding galaxy is conservatively classified as failing that criterion.
For H$\alpha$, we additionally estimate the continuum-subtracted signal within the feature bandpass and compute its signal-to-noise ratio using the SDSS flux uncertainties.
We note that though this method may not return precise line fluxes, its primary purpose to identify detectable emission or absorption features indicative of recent star formation and therefore provides a reliable EW-based diagnostic.
We find H:K ratios with {\tt pyLick} \citep[][]{Borghi22} as the ratio of the pseudo-equivalent widths of the features, computed using the standard Lick index definitions.
\begin{table}
    \centering
    \setlength{\tabcolsep}{4pt} 
    \caption{Bandpasses (start and width) for spectral features used to calculate equivalent widths.}
    \begin{tabular}{l|ccc}
        Index & Blue [\AA] & Feature [\AA] & Red [\AA] \\
        \hline
        [OII] & $3667.00+50.00$ & $3717.00+20.00$ & $3737.0+40.0$ \\
        H$\alpha$ & $6490.00+40.00$ & $6553.00+20.00$ & $6600.0+50.0$ \\
        H$\delta_\text{A}$ & $4041.60+38.15$ & $4083.50+38.75$ & $4128.5+32.5$ \\
    \end{tabular}
    \label{tab:bandpasses}
\end{table}

Following the recommendations of \citet[][ see also \citealt{Wang26}]{Moresco18}, we use the following prescriptions for each spectral feature:
\begin{itemize}
    \item No detectable [OII] emission: EW([OII])~$>-5$\AA
    \item No detectable H$\alpha$ emission: SNR(H$\alpha$)~$<3$
    \item No detectable H$\delta_\text{A}$\footnote{A-definition as opposed to F-definition for H$\delta$ where A is the ``wide'' $\sim40$\AA~definition and F is the ``narrow'' $\sim20$\AA ~definition. As H$\delta_\text{F}$ is narrower, it becomes harder to measure in galaxies with higher velocity dispersions \citep{Worthey97} the A-definition is more appropriate for our work.} absorption: EW(H$\delta_\text{A}$)~$<4$\AA
    \item Low contamination by young stellar population: H:K$ < 1$
\end{itemize}
Finally, we require galaxies to satisfy the commonly adopted stellar mass and velocity dispersion criteria, namely $\log(M_\star/M_\odot)>10.75$ and $\sigma_\star>250~\mathrm{km\,s^{-1}}$.
The stellar mass criterion is already satisfied by all galaxies in our final sample.

In total, out of the final relic sample, we find that only 30\% pass all of the CC selection criteria. 
We report the total number of relics that pass each individual test in Table~\ref{tab:passed}. 
The bottleneck is the low stellar velocity dispersion, with only 44\% of relics having $\sigma_{\star}>250$~km/s. 
However, $\sigma_{\star}$ cuts were designed to homogenise the sample in stellar mass.
There relation of $\sigma_{\star}$ vs M$_\star$ is not well defined in relics \citep[see][]{Mills25}, thus this not indicate that relics galaxies cannot be used as cosmic chronometers.
If we omit the velocity dispersion criterion, 87\% of galaxies pass the selection.
Overall, these results show that relic galaxies satisfy the vast majority of the traditional CC criteria. 
The principal difference arises from the velocity dispersion requirement, which excludes more than half of the relic sample. 
Consequently, although relic galaxies and classical CCs substantially overlap, the relic selection defines a considerably more restrictive subset of the passive galaxy population.

\begin{table}
    \centering
    \caption{Number and fraction of relic galaxies satisfying each of the traditional CC selection criteria. Applying all criteria simultaneously yields a final sample of 56 bona fide CC candidates.}
    \begin{tabular}{c|ccc}
        Criterion & \% Passed & Passed & Failed \\
        \hline
         UVJ & \phantom{0}96.8 & 183 & \phantom{00}6 \\
        $[\text{OII}]$ & \phantom{0}91.0 & 172 & \phantom{0}17 \\
        H$\alpha$ &  \phantom{0}98.4 & 186 & \phantom{00}3 \\
        H$\delta$ & 100.0 & 189 & \phantom{00}0 \\
        H:K & \phantom{0}87.8 & 166 & \phantom{0}23 \\
        M$_\star$ & 100.0 & 189 & \phantom{00}0 \\
        $\sigma_\star$ & \phantom{0}44.4 & \phantom{0}84 & 105 \\
        \hline
        Bona fide & \phantom{0}29.7 & \phantom{0}56 & 133 \\
    \end{tabular}
    \label{tab:passed}
\end{table}

\begin{figure}[ht]
\centering
\includegraphics[width = 0.48\textwidth]{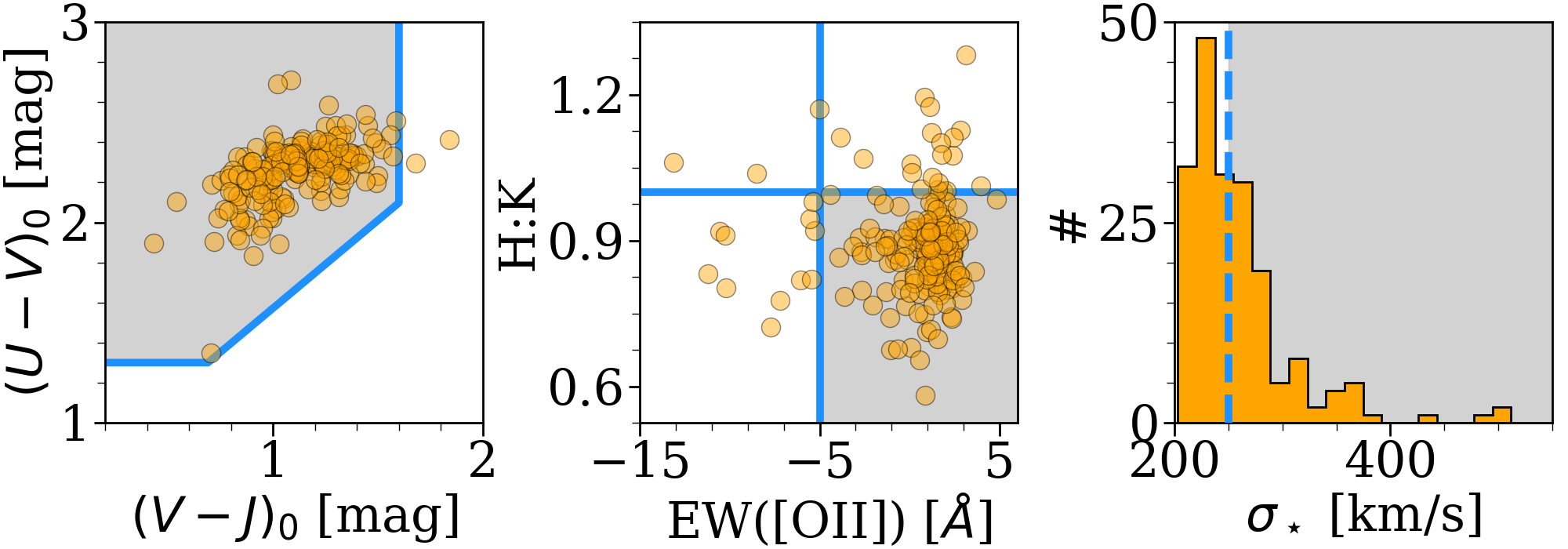}
\caption{Comparison of the final sample with traditional CC selection diagnostics. From left to right: $UVJ$ diagram, H:K ratio versus EW([OII]), and stellar velocity dispersion.}
\label{fig:traditionalCC}
\end{figure}

\section{The distribution of $A$}\label{app:Adistr}
To estimate the uncertainty on the normalisation $A(Z, [\alpha/\mathrm{Fe}])$, we perform a Monte Carlo propagation of the uncertainties in $Z$ and [$\alpha/\mathrm{Fe}]$. 
We generate 10\,000 realisations of the final sample by perturbing the median values of $Z$ and [$\alpha/\mathrm{Fe}]$ within their uncertainties (scaled MAD) assuming Gaussian distributions.
For each realisation, the corresponding value of $A$ is obtained by interpolating the $A(Z, [\alpha/\mathrm{Fe}])$ relation derived from the SPS models (Figure~\ref{fig:analysisA}). 
We restrict the sampling to the range covered by the models, $0 \leq [\alpha/\mathrm{Fe}] \leq 0.4$, and discard realisations falling outside this interval to avoid extrapolation beyond the calibrated parameter space.
The fraction of discarded realisations is $\sim$1\%.
Thus it would not significantly change our final error estimate.

In principle, the uncertainties in $Z$ and [$\alpha/\mathrm{Fe}]$ are expected to be correlated, both physically and through degeneracies in stellar population fitting. 
Consequently, treating them as independent Gaussian variables is an approximation. 
However, our final sample spans a very narrow metallicity range, resulting in a correspondingly narrow distribution of $A(Z)$ (Figure~\ref{fig:Ahist}). 
Quantifying the covariance between $Z$ and [$\alpha/\mathrm{Fe}]$ over such a limited metallicity range would require significantly more precise measurements of both quantities, together with SPS models sampling a wider range of [$\alpha/\mathrm{Fe}]$ values. 
As current models remain limited in this respect, we are unable to robustly quantify this covariance and therefore adopt the first-order approximation of independent uncertainties.

\begin{figure}[ht]
\centering
\includegraphics[width = 0.48\textwidth]{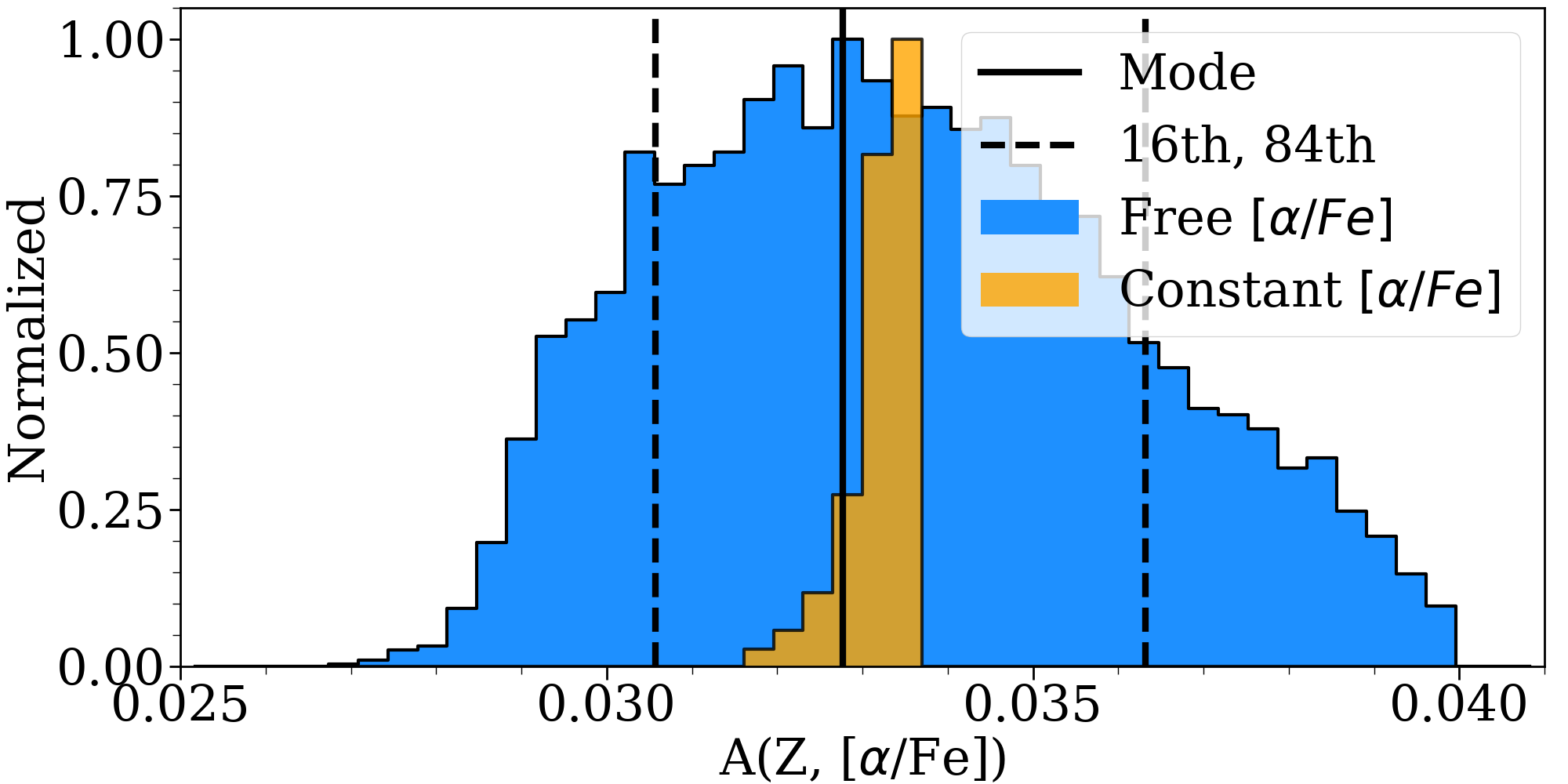}
\caption{Histogram of $A(Z, [\alpha/\textrm{Fe}]$) with propagated uncertainties. In orange, we present results of propagation only the $Z$ uncertainty, while in blue we propagate both $Z$ and $[\alpha/$Fe]. The black solid line represents the mode, while the black dashed lines show the 16th and 84th percentile of blue distribution.}
\label{fig:Ahist}
\end{figure}

\end{appendix}
\end{document}